\documentclass[sigconf]{acmart}
\AtBeginDocument{%
  }
\setcopyright{cc}
\setcctype{by-nc-nd}
\copyrightyear{2026}
\acmYear{2026}
\acmDOI{10.1145/3767308.3836100}
\acmConference[MM '26]{Proceedings of the 34th ACM International Conference on Multimedia}{November 10--14, 2026}{Rio de Janeiro, Brazil}
\acmBooktitle{Proceedings of the 34th ACM International Conference on Multimedia (MM '26), November 10--14, 2026, Rio de Janeiro, Brazil}
\acmISBN{979-8-4007-2213-4/2026/11}

\usepackage{booktabs}
\usepackage{multirow}
\usepackage{graphicx}
\usepackage{amsmath}
\usepackage{enumitem}
\usepackage{xspace}
\usepackage{cuted}
\usepackage{capt-of}
\usepackage{subcaption}
\usepackage{caption}
\usepackage{eso-pic}

\newcommand{\PlaceInstitutionLogos}{%
  \AddToShipoutPictureFG*{%
    \AtPageUpperLeft{%
      \raisebox{-1.40\height}{%
        \hspace*{0.20in}%
        \includegraphics[height=0.50in,trim=35 35 35 35,clip]{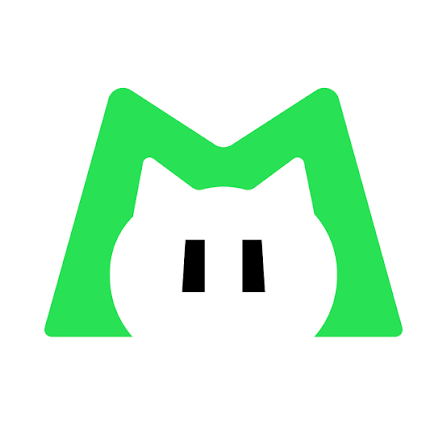}%
        \hspace{0.22in}%
        \includegraphics[height=0.44in]{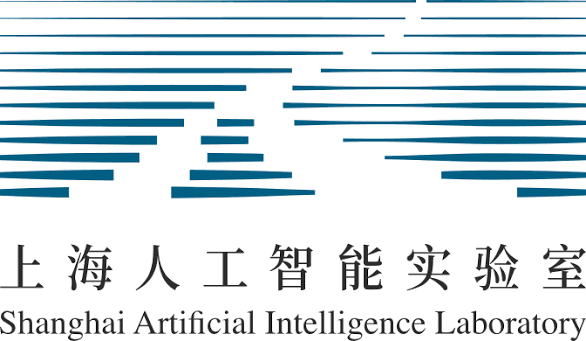}%
      }%
    }%
  }%
}

\newcommand{\ours}{PRISM-Bench}
\newcommand{\bench}{\ours{}}

\usepackage{array}
\usepackage{arydshln}
\usepackage{tikz}
\usepackage{array}
\usepackage{booktabs}
\usepackage{multirow}
\definecolor{SepGray}{gray}{0.78}
\newcommand{\sepv}{\color{SepGray}\vrule width 0.45pt}
\newcommand{\GTsep}{%
  \noalign{\vskip 0.2pt}
  \cline{2-24}
  \noalign{\vskip 0.8pt}
}
\newcommand{\NAblock}{/}
\newcommand{\best}[1]{\textbf{#1}}
\newcommand{\vsep}{\multicolumn{1}{@{\hspace{4pt}}c@{\hspace{4pt}}}{\sepv}}

\begin{document}
\PlaceInstitutionLogos
\settopmatter{printacmref=true,authorsperrow=4}
\title[\bench{}]{ PRISM-Bench: An Audio-Centric Diagnostic Benchmark for Text-to-Audio-Video Generation}

\author{Yuchen Sun}
\authornote{Yuchen Sun and Qian Yang contributed equally to this research. \newline This work was done while Yuchen Sun was an intern at Meituan.}
\email{sunyuchen@pjlab.org.cn}
\orcid{0009-0004-2672-0476}
\affiliation{%
  \institution{Shanghai Artificial Intelligence Laboratory}
  \city{Shanghai}\country{China}}

\author{Qian Yang}
\authornotemark[1]
\email{yangqian77@meituan.com}
\affiliation{%
  \institution{Meituan}
  \city{Beijing}
  \country{China}}

\author{Jun Wang}
\email{wangjun85@meituan.com}
\affiliation{%
  \institution{Meituan}
  \city{Beijing}
  \country{China}}

\author{Detai Xin}
\email{xindetai@meituan.com}
\orcid{0009-0007-1908-1137}
\affiliation{%
  \institution{Meituan}
  \city{Beijing}
  \country{China}}

\author{Guoqiao Yu}
\authornote{Corresponding authors: Guoqiao Yu and Guanglu Wan.}
\email{yuguoqiao@meituan.com}
\affiliation{%
  \institution{Meituan}
  \city{Beijing}
  \country{China}}

\author{Guanglu Wan}
\authornotemark[2]
\email{wanguanglu@meituan.com}
\affiliation{%
  \institution{Meituan}
  \city{Beijing}
  \country{China}}

\author{Qi Jia}
\email{jiaqi@pjlab.org.cn}
\orcid{0000-0001-6104-7249}
\affiliation{%
  \institution{Shanghai Artificial Intelligence Laboratory}
  \city{Shanghai}
  \country{China}}

\renewcommand{\shortauthors}{Yuchen Sun et al.}
 
\begin{abstract}
Text-to-audio-video (T2AV) generation has advanced rapidly, but its evaluation still underestimates the audio modality. Existing benchmarks either treat audio as an auxiliary component of video quality or assess it in isolation from audiovisual grounding, making it difficult to diagnose where current systems truly succeed or fail in audio generation. We present \bench{}, the first audio-centric diagnostic benchmark for T2AV generation. Built from a rigorously curated dataset of 900 human-verified samples, \bench{} factorizes audio evaluation along two orthogonal axes: audio type (Speech, Music, and Sound) and sound-source visibility (On-screen vs.\ Off-screen). It evaluates generated content across four perceptual dimensions (Audio-Visual Coherence, Audio Quality, Audio Expressiveness, and Prompt Following) with 35 fine-grained criteria. To ensure reliable assessment, we adopt an enhanced MLLM-as-a-Judge protocol based on blind, side-by-side comparison against ground-truth references, demonstrating strong alignment (over 70\% mean agreement) with human raters. Our evaluation of recent T2AV systems highlights a significant performance gap between frontier and open-source models. Furthermore, we demonstrate that current generation paradigms overfit to perceptual fidelity while struggling with complex grounding and control tasks, particularly in generating music and synchronized On-screen audio.\footnote{We provide a public leaderboard for reporting and comparing benchmark results at \href{https://huggingface.co/spaces/prismbench/prismbench-leaderboard}{\textcolor[HTML]{1A5FB4}{https://huggingface.co/spaces/prismbench/prismbench-leaderboard}}, while the underlying audiovisual data remain restricted due to licensing and redistribution constraints.}
\end{abstract}



\makeatletter
\let\prismSavedAuthorNotes\@authornotes
\let\@authornotes\@empty
\makeatother

\maketitle

\makeatletter
\begingroup
\setcounter{footnote}{0}
\renewcommand{\thefootnote}{\@fnsymbol\c@footnote}
\let\@footnotetext\@footnotetext@nolink
\prismSavedAuthorNotes
\endgroup
\setcounter{footnote}{0}
\makeatother

\section{Introduction}\label{sec:intro}
Text-to-audio-video (T2AV) generation has recently emerged as one of the most active frontiers in multimodal generative modeling. Recent systems now include a growing range of both proprietary and publicly released generators that produce video together with native or jointly modeled audio~\cite{openai2024sora,deepmind2024veo,kling2025omni,bytedance2026seedance2,hacohen2026ltx2,low2025ovi,openmoss2026mova}. Despite this rapid progress, evaluation has lagged noticeably behind. Recent benchmarks have expanded coverage to cross-modal alignment, physical plausibility, multiple generation tasks, and broad audio-video quality~\cite{jia2025vabench,sun2025t2avcompass,xie2025phyavbench,mao2024tavgbench}. However, their reported taxonomies are organized primarily around tasks, content scenarios, general evaluation dimensions, or physical failure modes, rather than a crossed diagnosis of both audio content type and sound-source visibility. As a result, they provide limited insight into how the same system's audio behavior changes jointly across speech, music, and sound and across On-screen and Off-screen sources.

\begin{figure*}[t]
\centering
\includegraphics
        [   width=\textwidth,
            trim=1 4 1 7,
            clip ]{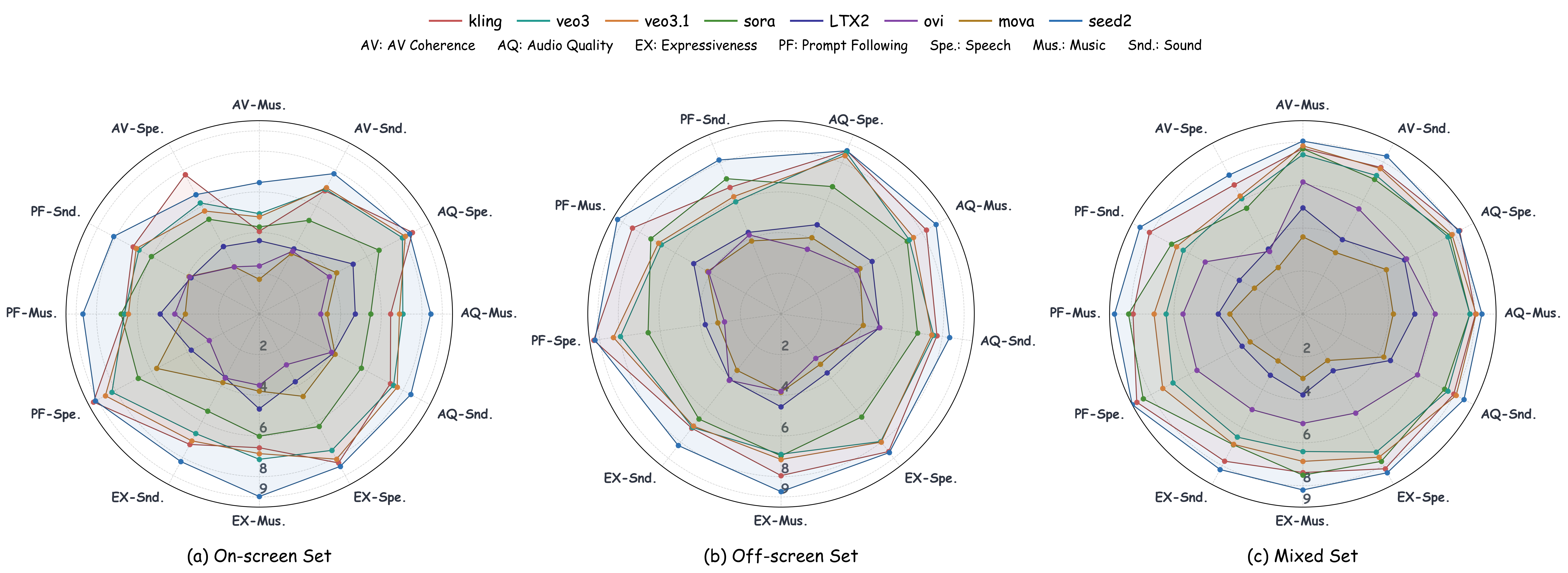}
\vspace{ -6 pt}
\caption{Grouped radar comparison across the On-screen, Off-screen, and Mixed subsets for the evaluated T2AV systems. Each spoke reports a calibrated mean score for one evaluation-dimension--audio-type pair; higher is better.}
\label{fig:grouped-radar}
\end{figure*}

This limitation is not a lack of fine-grained evaluation per se: several recent benchmarks define detailed dimensions or failure taxonomies~\cite{sun2025t2avcompass,jia2025vabench,xie2025phyavbench,zhou2026mtavgbench,guo2026alivebench,mao2024tavgbench}. The missing formulation is an explicit joint stratification over two factors that critically shape generation difficulty: the \emph{type of audio content} and the \emph{visibility of its source}. Evaluations spanning speech, music, and environmental sound have reported markedly different performance profiles across these domains~\cite{qiu2026audiocapbench}, while audio tied to visible events poses a different grounding problem from audio whose source remains off screen~\cite{tian2018ave,tzinis2022audioscopev2}. Without a crossed organization over these factors, aggregate results do not directly reveal where current T2AV systems succeed or fail under specific audio conditions.

\begin{figure*}[!t]
    \centering
    \begin{minipage}[t]{0.28\textwidth}
        \centering
        \includegraphics[
            pagebox=cropbox,
            width=\linewidth,
            trim=120 120 120 120,
            clip
        ]{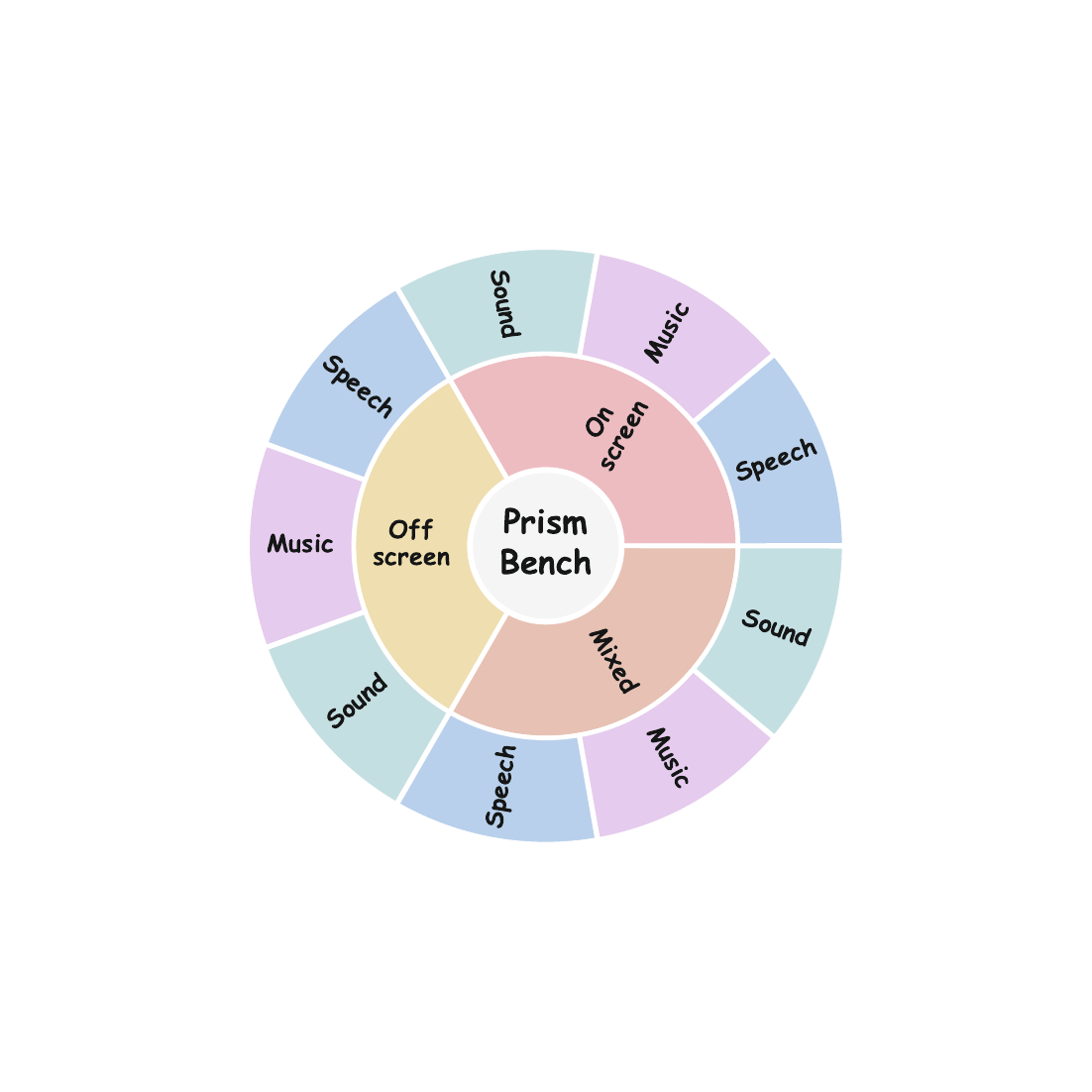}
        \vspace{0.15em}
        {\footnotesize (a) Factorized benchmark taxonomy}
    \end{minipage}
    \hspace{0.03\textwidth}
    \begin{minipage}[t]{0.30\textwidth}
        \centering
        \includegraphics[
            pagebox=cropbox,
            width=\linewidth,
            height=0.23\textheight,
            trim=6 5 2 4,
            clip
        ]{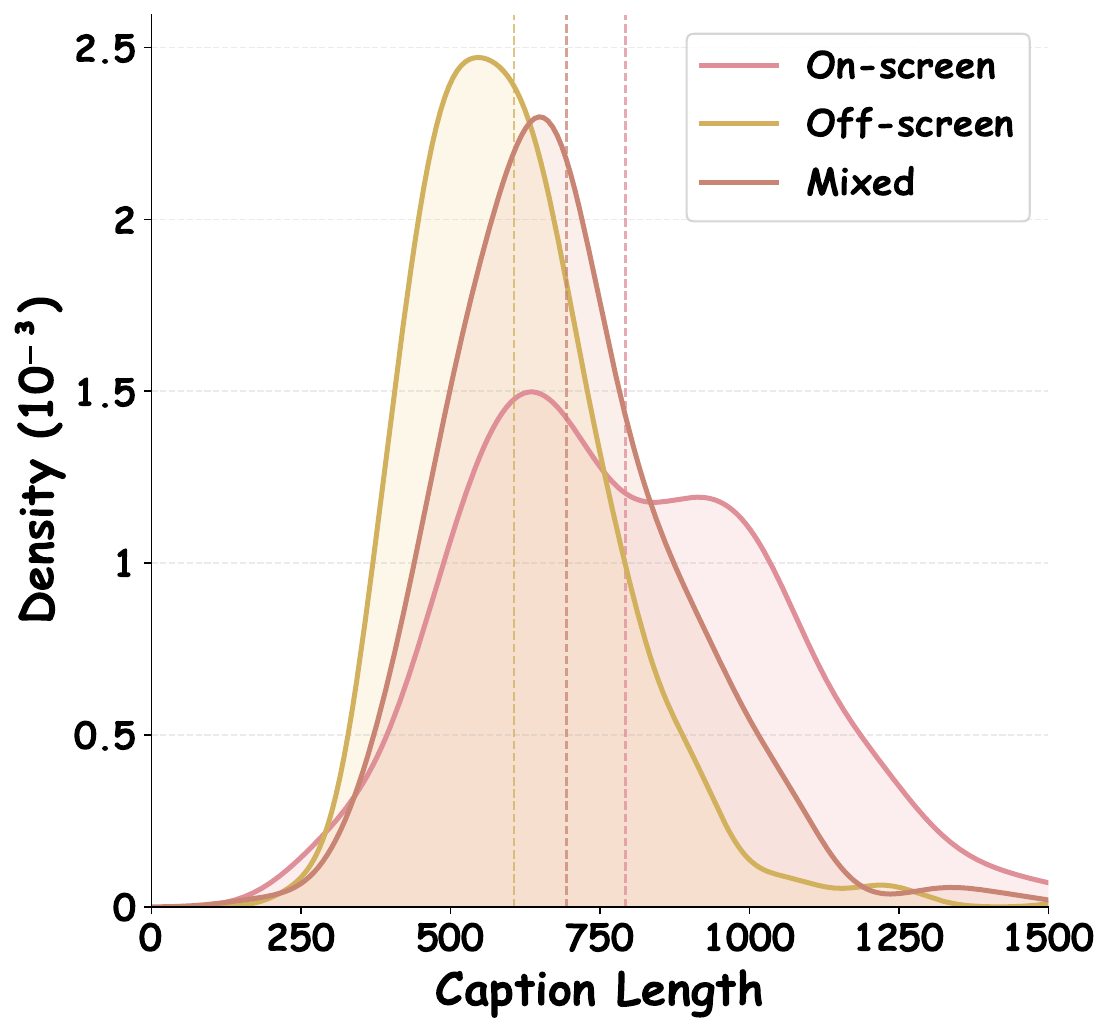}
        \vspace{0.15em}
        {\footnotesize (b) Caption-length distribution}
    \end{minipage}
    \hspace{0.03\textwidth}
    \begin{minipage}[t]{0.28\textwidth}
        \centering
        \includegraphics[
            pagebox=cropbox,
            width=\linewidth,
            trim=1 1 1 1,
            clip
        ]{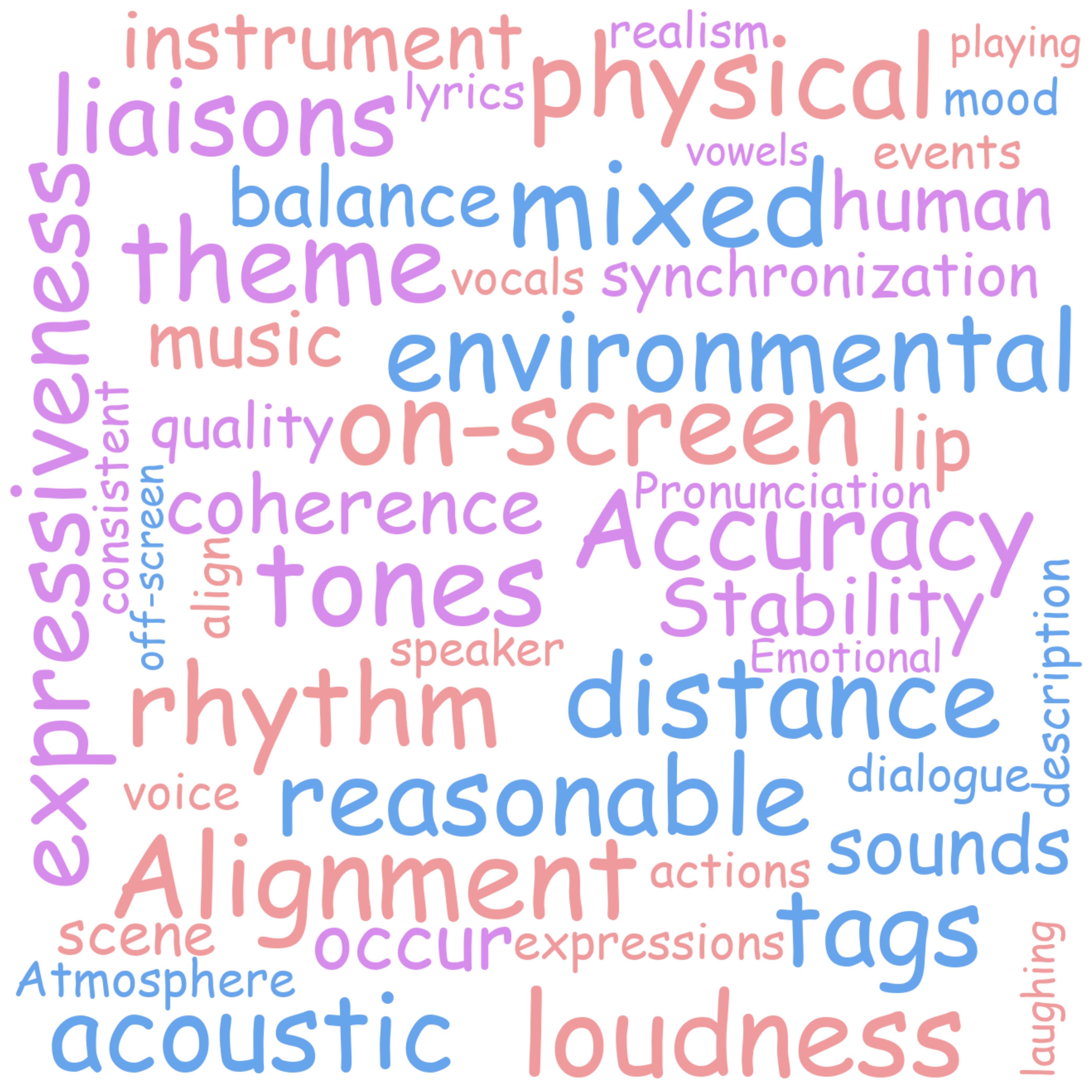}
        \vspace{0.15em}
        {\footnotesize (c) Eval Criteria Word Cloud}
    \end{minipage}
    \vspace{-6pt}
    \caption{Overall visualization of PRISM-Bench.}
    \label{fig:main}
\end{figure*}

To address these limitations, we introduce \bench{} (Figure~\ref{fig:main}), the first audio-centric diagnostic benchmark for T2AV generation. Rather than relying on holistic audio assessment, \bench{} explicitly models audio variation along two orthogonal axes: audio type, comprising Speech, Music, and Sound, and sound-source visibility, distinguishing On-screen from Off-screen audio.  \bench{} is built on a rigorously curated dataset of 900 human-verified samples, and evaluates generated content along four perceptual dimensions: Audio-Visual Coherence, Audio Quality, Audio Expressiveness, and Prompt Following. To make these dimensions operational for fine-grained diagnosis, we further instantiate them with 35 criteria that specify what the evaluator should examine under different audio types and visibility conditions.

Recent work has explored MLLMs as evaluators for image safety, audio-language responses, and any-to-any multimodal tasks~\cite{wang2025mllm,yang2024air,pu2025judge}. We design an enhanced side-by-side protocol anchored by ground-truth references. In each comparison, the judge is presented with a generated sample and a ground-truth sample in blinded and randomized order, and evaluates each candidate independently rather than through direct preference alone. We validate the framework through stability analysis and human-alignment experiments, showing that judge-side variation remains substantially smaller than the performance differences across candidate models, while the scores achieve over 70\% mean agreement with human raters. These results support the use of the protocol for large-scale benchmark evaluation.

Using \bench{}, we evaluate Seedance~2.0, Kling v3 Omni, Veo~3.0/3.1, Sora~2, LTX-2, Ovi, and MOVA, and find clear differences across models, audio types, visibility conditions, and evaluation dimensions (Figure~\ref{fig:grouped-radar}). Seedance~2.0 ranks first across the On-screen, Off-screen, and Mixed subsets, yet the factorized scores expose residual weaknesses that its aggregate lead would hide: On-screen AV Coherence remains lower than its fidelity- and control-oriented dimensions, and Off-screen Sound Prompt Following trails its Speech and Music results. Frontier proprietary systems also retain a clear advantage over open-source models. These findings indicate that improved acoustic realism and prompt adherence do not by themselves resolve fine-grained audiovisual grounding and sound-event control.

Our contributions are threefold:
\begin{itemize}[leftmargin=*,nosep]
\item We introduce \bench{}, audio-centric diagnostic benchmark for T2AV generation. By jointly stratifying audio type and sound-source visibility, \bench{} enables structured analysis under diverse grounding conditions.

\item We construct a rigorously curated dataset of 900 human-verified samples and pair it with a robust MLLM-as-a-Judge evaluation framework covering four perceptual dimensions and 35 fine-grained criteria.
\item We benchmark recent systems and provide a factorized empirical analysis of their capability profiles. Seedance~2.0 sets a new overall lead, while the remaining gaps in visible-source coherence and fine-grained Sound control demonstrate that aggregate progress does not eliminate grounding-sensitive failure modes.

\end{itemize}

We publicly release the \bench{} leaderboard and evaluation code at \url{https://huggingface.co/spaces/prismbench/prismbench-leaderboard}; the underlying audiovisual data remain restricted because redistribution rights vary across sources.

\section{Related Work}\label{sec:related}

\noindent\textbf{Text-to-Audio-Video Generation }Text-to-audio-video (T2AV) generation has progressed from loosely coupled pipelines to increasingly unified multimodal generators.

Early joint approaches such as MM-Diffusion modeled audio and video within a shared diffusion framework~\cite{ruan2023mmaking}.
More recent approaches scale joint generation with tighter cross-modal coupling, improving temporal alignment and semantic consistency between visual events and audio tracks.
Recent systems such as Sora~2~\cite{openai2024sora}, Veo~\cite{deepmind2024veo}, Movie Gen~\cite{polyak2024moviegen}, Kling v3 Omni~\cite{kling2025omni}, and SkyReels V4~\cite{chen2026skyreelsv4} increasingly target generation of temporally aligned video and audio. Seedance~2.0 adopts a unified multimodal audio-video generation architecture and adds dual-channel audio output for speech, music, and sound effects~\cite{bytedance2026seedance2}, illustrating how rapidly native audio capabilities are advancing.
At the same time, LTX-2~\cite{hacohen2026ltx2}, Ovi~\cite{low2025ovi}, and MOVA~\cite{openmoss2026mova} release model artifacts for reproducible study, making systematic comparison more feasible.
Despite this rapid progress, the audio side of T2AV generation remains notably less mature than the visual side.
Current models often exhibit uneven behavior across different audio categories, strong ambient sound generation but weak speech synchronization, or plausible music quality but poor event-level grounding.
Conventional automatic measures instead tend to target video distribution or motion~\cite{unterthiner2018fvd,liu2024fvmd,ge2024contentbiasfvd,luo2024beyondfvd}, audio distribution~\cite{kilgour2018fad}, or learned cross-modal similarity~\cite{elizalde2022clap,guzhov2021audioclip}; these scores do not by themselves provide a crossed diagnosis by audio type and source visibility.

\noindent\textbf{Audio-Visual Evaluation Benchmarks}
Audio-visual evaluation predates generative T2AV benchmarks and has traditionally focused on perceptual interactions between audio and video streams~\cite{min2020study}. However, evaluation for modern generative T2AV remains fragmented. Unimodal benchmarks such as VBench~\cite{huang2024vbench} and EvalCrafter~\cite{liu2024evalcrafter} assess general video quality, whereas AudioCaps~\cite{kim2019audiocaps} and related audio benchmarks focus on audio semantics, captioning, or fidelity. Although informative, these benchmarks do not directly evaluate whether generated audio is temporally coordinated with video, grounded in visible or implied events, and consistent with multimodal prompts.

Recent work has moved toward joint audiovisual evaluation. TAVGBench~\cite{mao2024tavgbench} introduces a large-scale text-to-audible-video dataset and AVHScore for audiovisual alignment, but does not jointly diagnose audio types and sound-source visibility conditions. T2AV-Compass~\cite{sun2025t2avcompass} combines signal-level metrics with MLLM-based judging over taxonomy-driven prompts, while VABench~\cite{jia2025vabench} evaluates multiple generation settings using comprehensive rubrics. PhyAVBench~\cite{xie2025phyavbench} focuses on physically grounded audiovisual events, MTAVG-Bench~\cite{zhou2026mtavgbench} targets multi-speaker conversational scenarios, and ALIVE~\cite{guo2026alivebench} evaluates 264 diverse prompts through blinded, side-by-side human comparisons across six categories. Complementary studies of LLM-as-a-Judge further reveal limitations of automatic evaluation, including presentation-order bias in pairwise judgments and the loss of uncertainty when human evaluation distributions are reduced to single-point scores~\cite{zheng2023llmjudge,verga2025distributional}.

Despite these advances, existing T2AV benchmarks do not provide a dedicated crossed diagnosis over sound type and source visibility. Consequently, they offer limited insight into how models behave across speech, music, and sound under On-screen and Off-screen conditions.


\section{PRISM-Bench}\label{sec:method}
\subsection{Overview}\label{sec:overview}

\bench{} is an audio-centric diagnostic benchmark for text-to-audio-video (T2AV) generation. It is designed to evaluate audio not as an auxiliary byproduct of video generation, but as a structured modality whose behavior depends on both audio type and grounding condition. To this end, \bench{} consists of two tightly coupled components: a curated benchmark dataset and a visibility-aware evaluation framework.

As illustrated in Figure~\ref{fig:main}(a), at the core of \bench{} is a factorized benchmark taxonomy organized along two axes: audio type and sound-source visibility. Their combination defines three complementary evaluation settings, namely On-screen, Off-screen, and Mixed, which together support more fine-grained diagnosis of model behavior than holistic audio-video evaluation. Built on top of this benchmark taxonomy, the evaluation framework assesses generated samples against ground-truth references along four perceptual dimensions: Audio-Visual Coherence, Audio Quality, Audio Expressiveness, and Prompt Following.


\subsection{Dataset Curation Pipeline}\label{sec:dataset_construction}
\begin{figure*}[t]
    \centering
    \includegraphics[width=0.97\textwidth,trim=3 126 10 101,clip]{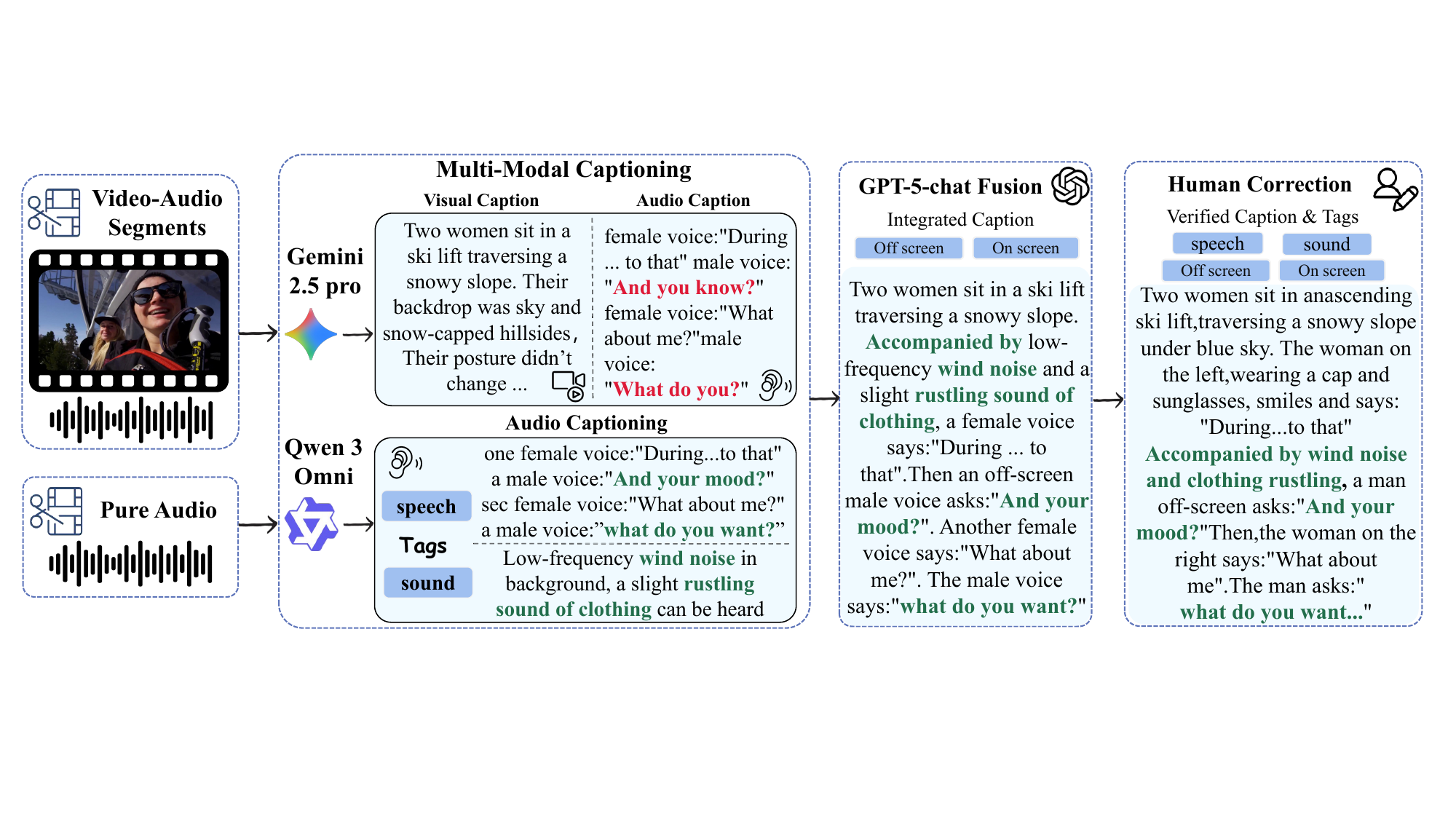}
    \vspace{ -5 pt}
    \caption{Data construction and annotation pipeline. }
    \label{fig:data-pipeline}
\end{figure*}

As illustrated in Figure~\ref{fig:data-pipeline}, we construct the dataset of \bench{} through a benchmark-oriented pipeline that integrates scenario-aware sampling, multi-stage annotation, and human verification. The goal is to support the two core diagnostic axes of \bench{}: audio type (\emph{Speech}, \emph{Music}, and \emph{Sound}) and sound-source visibility (\emph{On-screen} versus \emph{Off-screen}). Rather than collecting clips only for generic audio-visual diversity, we explicitly build the dataset to provide broad and balanced coverage of these benchmark-critical conditions.


The source pool comprises publicly accessible audiovisual content collected from web sources, together with audiovisual assets commercially licensed or purchased by our institution for internal research and evaluation. Because redistribution rights vary, the clips are retained in a controlled internal environment and are not publicly released. We instead provide a public leaderboard, with the data-access policy documented on the project page. 
To ensure ecological diversity, PySceneDetect\footnote{\url{https://github.com/breakthrough/pyscenedetect}} segments source videos into candidate scenes, which we balance across Movie, Dialogue-centric, Documentary, Concert, and Daily-life content while monitoring \emph{Speech}, \emph{Music}, \emph{Sound}, \emph{On-screen}, and \emph{Off-screen} coverage.

After sampling, all videos are divided into clips of up to $8$ seconds through the Video-Audio Segments step, which defines the basic evaluation unit of the benchmark. This granularity preserves meaningful audio-visual events while keeping each sample compact enough for controlled evaluation. We then apply a multi-stage annotation pipeline that combines multimodal captioning, audio-only calibration, cross-source fusion, and benchmark-level human correction.

In the first annotation stage, we apply \textbf{Multi-Modal Captioning} using the stable API model \textbf{Gemini 2.5 Pro}\footnote{\url{https://ai.google.dev/gemini-api/docs/models/gemini-2.5-pro}} to obtain an initial multimodal description for each clip. Specifically, Gemini produces two complementary captions: a visual caption describing scenes, objects, and actions, and an audio caption describing audible content such as speech, music, and sound events. These captions provide broad multimodal coverage and serve as the basis for subsequent refinement.

In the second stage, we perform \textbf{Audio Captioning} by extracting the raw audio track from each clip and feeding it to \textbf{Qwen3-Omni}~\cite{xu2025qwen3omni}. We use Qwen3-Omni in an audio-only setting, and empirically find that pure-audio input produces more accurate audio captions than audio-video input, allowing the model to focus more directly on acoustic evidence. In addition to audio descriptions, Qwen3-Omni also produces auxiliary attributes such as language information that are later used for filtering and analysis. This stage is particularly useful for improving audio-centric annotation when multimodal perception alone is insufficient.

In the third stage, we conduct \textbf{GPT-5 Chat Fusion} using the API alias gpt-5-chat-latest, which the provider documents as pointing to the dated snapshot gpt-5-chat-latest-2025-08-07~\cite{openai2025gpt5systemcard}. The model takes as input both the multimodal captions from Gemini and the audio-only descriptions from \textbf{Qwen3-Omni}, and reconciles them into a more consistent clip-level annotation. This stage improves cross-source consistency, reduces hallucinated audio-visual correspondences, and refines incomplete or ambiguous descriptions. More importantly, it introduces the structured labels required by \bench{}: the fusion model identifies which audio types are present in each clip among \emph{Speech}, \emph{Music}, and \emph{Sound}, and determines whether the relevant sound source is \emph{On-screen} or \emph{Off-screen}.

Finally, all 900 benchmark samples undergo \textbf{Human Correction}, where annotators verify benchmark-critical attributes, including synchronization descriptions, speech content, paralinguistic cues, music and environmental sound events, and the labels of audio type and sound-source visibility. This final quality-control step makes the annotations suitable for fine-grained diagnostic evaluation. The benchmark contains three visibility-aware sets of 300 audio-video pairs each.

The benchmark annotations also exhibit clear regularities. As shown in Figure~\ref{fig:main}(b), the caption-length distributions differ systematically across the three benchmark settings, reflecting different descriptive demands under On-screen, Off-screen, and Mixed conditions. 
On-screen samples tend to have longer captions because they require joint description of audible content, visible sources, and their grounding relations, whereas Off-screen samples are generally shorter since they do not require explicit visible-source alignment. Mixed samples lie between these two cases.

\begin{figure*}[!t]
    \centering
    \includegraphics[
        width=0.92\textwidth,
        trim=5 95 10 95,
        clip
    ]{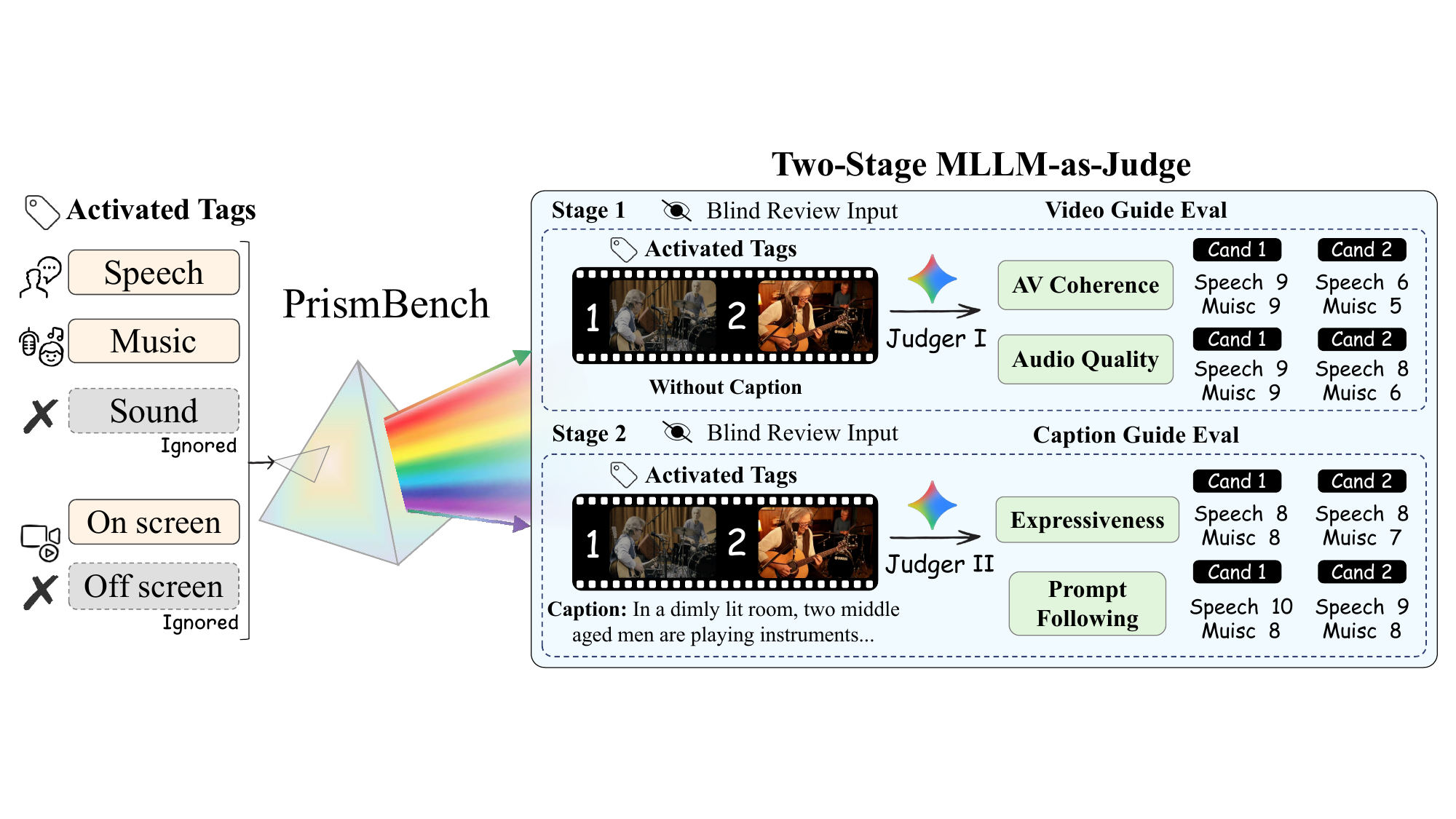}
    \vspace{ -6 pt}
    \caption{PRISM-Bench evaluation pipeline.}
    \label{fig:prism-bench}
\end{figure*}

\subsection{Evaluation Dimensions}\label{sec:eval_dim}

\bench{} evaluates T2AV outputs along four complementary perceptual dimensions, which together capture grounding, fidelity, expressiveness, and instruction adherence in audio generation. Rather than applying the same criteria uniformly to all samples, the benchmark instantiates each dimension with type-aware criteria for Speech, Music, and Sound, and activates them conditionally according to the sample's activated tags and visibility condition. A lexical overview of these criteria is shown in Figure~\ref{fig:main}(c). This design allows the evaluation to remain both diagnostic and comparable across heterogeneous audio scenarios.

\noindent\textbf{Audio-Visual Coherence}
 measures whether generated audio is properly grounded in visible events and sounding sources, following the broader audio-visual alignment and on-screen grounding problems studied in prior work~\cite{mao2024tavgbench,mo2024t2avbench,tian2018ave,tzinis2022audioscopev2}. It is activated only when the relevant audio condition is visually grounded, and therefore applies to On-screen and the On-screen portion of Mixed samples, but remains inactive for purely Off-screen cases. This dimension focuses on temporal alignment and physical plausibility, including whether speech is synchronized with visible speaking, whether musical sounds correspond to visible playing actions, and whether sound events are consistent with visible sources and their timing.

\noindent\textbf{Audio Quality}
evaluates the perceptual fidelity and technical realism of the audio signal itself. For Speech, it concerns naturalness, clarity, and vocal realism. For Music, it captures properties such as timbral plausibility, rhythmic stability, and overall musical coherence. For Sound, it focuses on whether the generated audio is physically believable, acoustically well-formed, and free from obvious synthetic artifacts or degradation. This dimension is intended to measure how realistic the audio sounds, independent of whether it is well grounded or well controlled.

\noindent\textbf{Audio Expressiveness}
assesses whether the generated audio conveys appropriate emotion, atmosphere, and scene-level support beyond basic fidelity. For Speech, this includes the adequacy of speaking style, emotional tone, and delivery. For Music, it concerns whether the music appropriately conveys mood and expressive character. For Sound, it reflects whether the audio contributes plausibly to scene atmosphere and perceptual vividness. 


\noindent\textbf{Prompt Following}
measures how faithfully the generated audio satisfies the textual instruction. This dimension is restricted to auditory adherence, so that the benchmark remains aligned with its audio-centric scope. For Speech, it considers verbal content and vocal characteristics; for Music, it concerns style, instrumentation, and mood; and for Sound, it focuses on event accuracy and texture correspondence. By isolating audio-side compliance from unrelated visual factors, this dimension makes the final scores more directly interpretable for T2AV evaluation.

\subsection{Evaluation Framework}\label{sec:eval_framework}


As illustrated in Figure~\ref{fig:prism-bench}, \bench{} adopts a tag-conditioned and visibility-aware evaluation framework for audio-centric T2AV assessment. Given a generated sample and its ground-truth reference, the framework activates only the evaluation targets relevant to the sample's annotated audio conditions, including both the applicable dimensions and the corresponding type-specific rubrics, and evaluates them under a fixed blind judging protocol. This design enables \bench{} to support not only overall model comparison, but also fine-grained and interpretable diagnosis across heterogeneous audio scenarios.

\noindent \textbf{Tag-Conditioned Evaluation.}
Using the human-verified activated tags defined in Section~\ref{sec:dataset_construction}, \bench{} partitions evaluation into three settings: On-screen, Off-screen, and Mixed. These tags specify which audio types are present in a sample and whether the relevant sound sources are visually grounded. During evaluation, only the dimensions and rubrics selected by the activated tags are included in the judge input. As example shown in Figure~\ref{fig:prism-bench}, the sample activates \emph{Speech} and \emph{On-screen}, while \emph{Sound} and \emph{Off-screen} are ignored. Accordingly, Stage~1 evaluates only the activated criteria for AV Coherence and Audio Quality from the merged clip without caption, whereas Stage~2 evaluates Speech-related Expressiveness and Prompt Following with access to the caption. This conditional activation mechanism keeps the evaluation focused on the relevant audio conditions and avoids introducing irrelevant scoring criteria into the judging process.


\noindent\textbf{Blind Side-by-Side Review.}
To ensure direct comparability under a shared perceptual context, we adopt a blind side-by-side review protocol. As shown in Figure~\ref{fig:prism-bench}, the evaluation model receives two candidate videos as separate inputs. In our implementation, each video begins with a 0.5-second black identifier frame labeled ``1'' or ``2'', which is used only to map the outputs to Candidate~1 and Candidate~2 and is explicitly ignored during evaluation. The model then evaluates the two candidates under matched viewing conditions without being told which one is the generated sample and which one is the ground-truth reference. This design reduces presentation bias and preserves a controlled comparison setting.

\noindent\textbf{Two-Stage Judge Design.}
We use Gemini-3.1-pro-preview~\cite{gemini31propreview2026}\footnote{Official Doc:  \url{https://ai.google.dev/gemini-api/docs/models/gemini-3.1-pro-preview}.} as the judge model and keep its nominal decoding configuration fixed. We set the temperature to 0.0 to remove sampling temperature as a varying factor, while recognizing that this setting does not guarantee identical outputs across repeated hosted-API calls: backend execution and numerical effects can still introduce variation even under greedy decoding~\cite{yuan2025fp32,li2026temperaturejudge,atil2025nondeterminism,messina2025backgroundtemperature}. To align each request with the evidence required by the target dimensions, the evaluation is divided into two stages, as illustrated in Figure~\ref{fig:prism-bench}. The first stage evaluates Audio-Visual Coherence and Audio Quality from the merged clip together with each sample's activated tags, which specify the audio categories to be evaluated. The second stage evaluates Audio Expressiveness and Prompt Following with access to the human-verified caption, which serves as the reference prompt and corresponds to the text prompt used to generate the tested video. This two-stage design separates perception-driven evaluation from prompt-conditioned evaluation, thereby improving the clarity and consistency of the judging process.

\noindent\textbf{Absolute Scoring with Evidence.}
Within each stage, the judge scores Candidate~1 and Candidate~2 independently against fixed rubric descriptions, rather than assigning only a relative preference. This is a pointwise, rubric-based design consistent with form-filling and reliability-oriented LLM-evaluation frameworks~\cite{liu2023geval,gu2024llmasjudge_survey}. In addition, the judge produces a brief evidence statement for each active dimension. These evidence traces make the evaluation process more transparent, support later auditing and error analysis, and provide qualitative support for understanding model strengths and weaknesses beyond the numerical scores alone.

\section{Experiments}\label{sec:exp}
\subsection{T2AV Generation Models}\label{sec:setup}

We evaluate the following recent text-to-audio-video generation systems on \bench{}.

\noindent\textbf{Seedance~2.0}~\cite{bytedance2026seedance2}\footnote{\url{https://seed.bytedance.com/seedance2_0}.}. Seedance~2.0 uses a unified multimodal audio-video generation architecture with native dual-channel audio. We evaluate its text-conditioned audiovisual outputs on the same benchmark prompts used for the other systems.

\noindent\textbf{Kling v3 Omni}~\cite{kling2025omni}. We access Kling v3 Omni through the official API\footnote{\url{https://kling.ai/document-api/apiReference/model/OmniVideo}.} and evaluate the native audiovisual outputs returned by the system without additional post-processing.

\noindent\textbf{Veo~3.0 and Veo~3.1}~\cite{deepmind2024veo}. We access both versions through the official Gemini API\footnote{\url{https://ai.google.dev/gemini-api/docs/video}.} and evaluate synchronized audiovisual clips generated at $1280 \times 720$ resolution with an 8-second duration.

\noindent\textbf{Sora~2}~\cite{openai2024sora}. We access Sora~2 through the official OpenAI video API\footnote{\url{https://developers.openai.com/api/docs/guides/video-generation}.} and evaluate native system outputs generated at $1280 \times 720$ resolution with an 8-second duration.

\noindent\textbf{LTX-2}~\cite{hacohen2026ltx2}. We run LTX-2 locally with the official repository\footnote{\url{https://github.com/Lightricks/LTX-2}.}~\cite{hacohen2026ltx2} and evaluate outputs generated with the released LTX-2 pipeline. Our setup uses the 19B development model together with the official upsampling pipeline, producing 8-second clips at $704 \times 1280$ and 24 FPS with 40 sampling steps and classifier-free guidance scale 4.0.

\noindent\textbf{Ovi}~\cite{low2025ovi}. We evaluate Ovi locally using the official open-source release\footnote{\url{https://github.com/character-ai/Ovi}.}. We adopt the official text-to-video generation pipeline and evaluate outputs at $720 \times 1280$ resolution with a 10-second duration, using UniPC sampling with 50 steps, audio guidance scale 3.0, and video guidance scale 4.0.

\noindent\textbf{MOVA}~\cite{openmoss2026mova}. We evaluate MOVA locally using the public codebase\footnote{\url{https://github.com/OpenMOSS/MOVA}}. We follow the official T2VA testing protocol and use a pure white image as the first-frame condition for text-driven generation; in our preliminary experiments, this mode occasionally produced corrupted frames or black-screen failures. We evaluate outputs at $720 \times 1280$ resolution with a 10-second duration, using the default official inference path with classifier-free guidance scale 5.0.

\subsection{PRISM-Bench Evaluation Results}\label{sec:main_results}

To mitigate context-dependent variation in ground-truth assessments, the mean model score for each metric-by-audio-type cell is normalized by the corresponding paired ground-truth mean and rescaled to the global ground-truth mean. The Final score is computed as the sum of all active calibrated cells.

The evaluation results of representative recent T2AV systems are summarized in Table~\ref{tab:main_results_all}. Following the visibility-aware design of \bench{}, we report results separately on the three benchmark subsets, namely On-screen, Off-screen, and Mixed. Within each subset, performance is evaluated along four perceptual dimensions, and is further broken down by the three audio types. This layout allows the table to present model behavior simultaneously from three complementary perspectives: evaluation dimension, audio type, and visibility condition. 
As a result, the table supports a more diagnostic view of model capability than holistic evaluation alone.
The following discussion provides a detailed characterization of these observations, along with their model-level implications and insights.

\noindent\textbf{Frontier Proprietary Systems Maintain a Clear Advantage.}

\begin{table*}[t]
  \centering
  \scriptsize
  \setlength{\tabcolsep}{3pt}
  \renewcommand{\arraystretch}{1.10}
  \caption{Results across the three Subsets, reported as GT-calibrated means on a 0--10 scale. Within each dimension, Tot.\ sums Speech (Spe.), Music (Mus.), and Sound (Snd.); Final sums all applicable totals. AV Coherence is marked as '/' when Off-screen.
  }
  \label{tab:main_results_all}
  \vspace{-4 pt}
  \resizebox{0.98\textwidth}{!}{%
  \begin{tabular}{
    @{}
    >{\centering\arraybackslash}m{1.0cm}
    >{\centering\arraybackslash}m{1.6cm}
    c
    cccc
    cccc
    cccc
    cccc
    c
    ccc
    c
    @{}
  }
    \toprule
    \multicolumn{1}{c}{} &
    \multicolumn{1}{c}{} &
    \multicolumn{1}{c}{} &
    \multicolumn{4}{c}{\textbf{Audio Quality}} &
    \multicolumn{4}{c}{\textbf{Audio Expressiveness}} &
    \multicolumn{4}{c}{\textbf{Prompt Following}} &
    \multicolumn{4}{c}{\textbf{AV Coherence}} &
    \multicolumn{1}{c}{} &
    \multicolumn{3}{c}{\textbf{Total}} &
    \multicolumn{1}{c}{} \\
    \cmidrule(lr){4-7}\cmidrule(lr){8-11}\cmidrule(lr){12-15}\cmidrule(lr){16-19}\cmidrule(lr){21-23}
    \multicolumn{1}{c}{\textbf{Subset}} &
    \multicolumn{1}{c}{\textbf{Model}} &
    \multicolumn{1}{c}{} &
    Spe. & Mus. & Snd. & \textbf{Tot.} &
    Spe. & Mus. & Snd. & \textbf{Tot.} &
    Spe. & Mus. & Snd. & \textbf{Tot.} &
    Spe. & Mus. & Snd. & \textbf{Tot.} &
    \multicolumn{1}{c}{} &
    Spe. & Mus. & Snd. &
    \textbf{Final} \\
    \midrule

    \multirow{9}{*}{\centering On-screen}
      & Ground Truth & \vsep & 8.06 & 7.54 & 7.50 & 23.10 & 8.00 & 7.62 & 7.32 & 22.94 & 8.62 & 8.24 & 7.95 & 24.81 & 7.23 & 6.53 & 7.17 & 20.93 & \vsep & 31.91 & 29.94 & 29.93 & 91.78 \\
      \GTsep
      & Seedance~2.0 & \vsep & 8.37 & \best{8.43} & \best{8.44} & \best{25.24} & \best{8.49} & \best{8.98} & \best{8.22} & \best{25.69} & 9.12 & \best{8.67} & \best{8.11} & \best{25.89} & 6.64 & \best{6.46} & \best{7.81} & \best{20.91} & \vsep & 32.62 & \best{32.54} & \best{32.58} & \best{97.73} \\
      & Kling-v3-omni & \vsep & \best{8.53} & 6.46 & 7.30 & 22.29 & 8.30 & 6.58 & 7.27 & 22.14 & \best{9.23} & 6.62 & 7.03 & 22.89 & \best{7.75} & 4.06 & 6.86 & 18.67 & \vsep & \best{33.82} & 23.72 & 28.46 & 85.99 \\
      & Veo~3.1 & \vsep & 8.13 & 6.89 & 7.68 & 22.70 & 8.09 & 6.87 & 7.07 & 22.03 & 8.57 & 6.43 & 6.85 & 21.85 & 5.73 & 4.77 & 7.03 & 17.54 & \vsep & 30.52 & 24.96 & 28.64 & 84.11 \\
      & Veo~3.0 & \vsep & 7.97 & 7.06 & 7.48 & 22.51 & 7.60 & 7.15 & 6.66 & 21.42 & 8.20 & 6.63 & 6.70 & 21.54 & 6.18 & 4.93 & 6.95 & 18.06 & \vsep & 29.96 & 25.77 & 27.79 & 83.52 \\
      & Sora~2 & \vsep & 6.67 & 5.48 & 5.68 & 17.83 & 6.27 & 6.01 & 5.41 & 17.69 & 6.74 & 6.79 & 6.01 & 19.55 & 5.27 & 4.28 & 5.21 & 14.76 & \vsep & 24.95 & 22.56 & 22.31 & 69.82 \\
      & LTX-2 & \vsep & 5.22 & 4.72 & 4.13 & 14.08 & 3.77 & 4.67 & 3.55 & 11.98 & 3.79 & 4.87 & 3.80 & 12.47 & 3.75 & 3.60 & 3.62 & 10.98 & \vsep & 16.53 & 17.87 & 15.10 & 49.50 \\
      & Ovi & \vsep & 3.90 & 3.02 & 4.02 & 10.95 & 2.83 & 3.51 & 3.57 & 9.92 & 2.78 & 4.16 & 3.88 & 10.82 & 2.62 & 2.36 & 3.49 & 8.47 & \vsep & 12.13 & 13.06 & 14.97 & 40.15 \\
      & MOVA & \vsep & 4.31 & 3.34 & 4.22 & 11.88 & 4.59 & 3.80 & 3.82 & 12.21 & 5.71 & 3.64 & 3.91 & 13.25 & 2.63 & 1.70 & 3.36 & 7.69 & \vsep & 17.24 & 12.48 & 15.31 & 45.04 \\
    \midrule

    \multirow{9}{*}{\centering Off-screen}
      & Ground Truth & \vsep & 8.19 & 7.86 & 7.56 & 23.60 & 8.08 & 7.65 & 7.16 & 22.89 & 8.76 & 8.28 & 7.56 & 24.61 & \multicolumn{4}{c}{\NAblock} & \vsep & 25.04 & 23.79 & 22.28 & 71.10 \\
      \GTsep
      & Seedance~2.0 & \vsep & 8.65 & \best{8.80} & \best{8.37} & \best{25.82} & \best{8.65} & \best{8.74} & \best{8.21} & \best{25.60} & 9.24 & \best{9.29} & \best{8.16} & \best{26.69} & \multicolumn{4}{c}{\NAblock} & \vsep & \best{26.54} & \best{26.83} & \best{24.74} & \best{78.11} \\
      & Kling-v3-omni & \vsep & \best{8.65} & 8.25 & 7.74 & 24.64 & 8.59 & 7.95 & 7.13 & 23.67 & \best{9.30} & 8.44 & 6.71 & 24.44 & \multicolumn{4}{c}{\NAblock} & \vsep & 26.54 & 24.63 & 21.59 & 72.76 \\
      & Veo~3.1 & \vsep & 8.39 & 7.52 & 7.48 & 23.38 & 8.00 & 7.16 & 7.00 & 22.16 & 8.32 & 6.96 & 6.22 & 21.49 & \multicolumn{4}{c}{\NAblock} & \vsep & 24.70 & 21.63 & 20.70 & 67.03 \\
      & Veo~3.0 & \vsep & 8.57 & 7.28 & 7.58 & 23.42 & 7.96 & 6.90 & 7.08 & 21.93 & 7.97 & 6.80 & 5.95 & 20.73 & \multicolumn{4}{c}{\NAblock} & \vsep & 24.50 & 20.97 & 20.61 & 66.08 \\
      & Sora~2 & \vsep & 6.75 & 7.16 & 6.78 & 20.70 & 6.43 & 6.98 & 6.56 & 19.97 & 6.60 & 7.39 & 7.17 & 21.16 & \multicolumn{4}{c}{\NAblock} & \vsep & 19.79 & 21.53 & 20.51 & 61.83 \\
      & LTX-2 & \vsep & 4.73 & 5.16 & 4.88 & 14.77 & 3.68 & 4.57 & 4.09 & 12.34 & 3.76 & 4.96 & 4.32 & 13.05 & \multicolumn{4}{c}{\NAblock} & \vsep & 12.17 & 14.69 & 13.29 & 40.16 \\
      & Ovi & \vsep & 3.43 & 4.30 & 4.90 & 12.62 & 2.77 & 3.81 & 4.14 & 10.72 & 2.81 & 4.11 & 4.19 & 11.11 & \multicolumn{4}{c}{\NAblock} & \vsep & 9.02 & 12.22 & 13.22 & 34.46 \\
      & MOVA & \vsep & 4.04 & 4.47 & 4.08 & 12.58 & 3.14 & 3.85 & 3.52 & 10.51 & 3.14 & 4.17 & 3.87 & 11.18 & \multicolumn{4}{c}{\NAblock} & \vsep & 10.32 & 12.49 & 11.47 & 34.28 \\
    \midrule

    \multirow{9}{*}{\centering Mixed}
      & Ground Truth & \vsep & 8.17 & 8.20 & 7.88 & 24.25 & 7.75 & 7.42 & 7.14 & 22.31 & 8.37 & 8.08 & 7.63 & 24.08 & 7.33 & 8.07 & 7.65 & 23.05 & \vsep & 31.62 & 31.77 & 30.30 & 93.69 \\
      \GTsep
      & Seedance~2.0 & \vsep & 8.22 & \best{8.34} & \best{8.50} & \best{25.06} & \best{8.38} & \best{8.20} & \best{8.21} & \best{24.78} & \best{9.00} & \best{8.77} & \best{8.59} & \best{26.35} & \best{7.32} & \best{8.05} & \best{8.32} & \best{23.69} & \vsep & \best{32.91} & \best{33.35} & \best{33.62} & \best{99.88} \\
      & Kling-v3-omni & \vsep & \best{8.27} & 8.04 & 7.96 & 24.27 & 8.17 & 7.39 & 7.77 & 23.32 & 8.76 & 7.91 & 8.09 & 24.76 & 6.81 & 7.71 & 7.73 & 22.26 & \vsep & 32.00 & 31.05 & 31.55 & 94.61 \\
      & Veo~3.1 & \vsep & 7.87 & 8.07 & 8.09 & 24.02 & 7.56 & 6.87 & 6.88 & 21.31 & 7.39 & 6.93 & 6.67 & 20.99 & 6.22 & 7.83 & 7.67 & 21.72 & \vsep & 29.03 & 29.70 & 29.30 & 88.04 \\
      & Veo~3.0 & \vsep & 7.65 & 7.78 & 7.65 & 23.08 & 7.29 & 6.41 & 6.50 & 20.20 & 6.85 & 6.37 & 6.31 & 19.53 & 6.08 & 7.42 & 7.30 & 20.79 & \vsep & 27.87 & 27.97 & 27.76 & 83.61 \\
      & Sora~2 & \vsep & 7.77 & 7.78 & 7.47 & 23.02 & 7.78 & 7.51 & 6.90 & 22.19 & 8.42 & 8.12 & 6.93 & 23.47 & 5.58 & 7.69 & 7.10 & 20.36 & \vsep & 29.55 & 31.10 & 28.39 & 89.04 \\
      & LTX-2 & \vsep & 5.36 & 5.21 & 4.62 & 15.19 & 3.00 & 3.77 & 3.23 & 10.00 & 3.21 & 3.94 & 3.35 & 10.49 & 3.42 & 4.94 & 3.92 & 12.29 & \vsep & 14.99 & 17.85 & 15.12 & 47.96 \\
      & Ovi & \vsep & 5.47 & 6.16 & 6.05 & 17.68 & 5.23 & 5.10 & 5.05 & 15.38 & 5.59 & 5.58 & 5.16 & 16.33 & 3.30 & 6.15 & 5.54 & 14.98 & \vsep & 19.59 & 22.99 & 21.80 & 64.38 \\
      & MOVA & \vsep & 4.40 & 4.21 & 4.27 & 12.88 & 2.46 & 3.00 & 2.48 & 7.94 & 2.78 & 3.40 & 2.56 & 8.75 & 2.45 & 3.59 & 3.24 & 9.28 & \vsep & 12.09 & 14.20 & 12.55 & 38.85 \\
    \bottomrule
  \end{tabular}%
  }
\end{table*}

Table~\ref{tab:main_results_all} shows a substantial and consistent gap between frontier proprietary systems and current open-source models across all three subsets. Seedance~2.0 ranks first on On-screen, Off-screen, and Mixed with Final scores of 97.73, 78.11, and 99.88, respectively; Kling-v3-omni is the runner-up on the first two subsets with 85.99 and 72.76, and on Mixed with 94.61. By comparison, the strongest open-source result is 49.50 on On-screen (LTX-2), 40.16 on Off-screen (LTX-2), and 64.38 on Mixed (Ovi). Thus, the lead of Seedance~2.0 over the strongest open model remains at least 35 points on every subset.
The dimension-wise results reveal where this advantage arises. On On-screen, Seedance~2.0 reaches 25.24 in Audio Quality, 25.69 in Audio Expressiveness, and 25.89 in Prompt Following, producing a Final score of 97.73 despite a lower AV Coherence total of 20.91. These results show that recent proprietary systems have advanced native audio generation substantially, while current open-source alternatives remain far behind across both fidelity- and grounding-sensitive dimensions.

\noindent\textbf{Audio Fidelity and Prompt Adherence Advance Faster than Visible-Source Grounding.}
Seedance~2.0 exceeds the ground-truth Audio Quality total on On-screen, Off-screen, and Mixed. It also exceeds the corresponding ground-truth Prompt Following total on all three subsets. These numbers indicate substantial progress not only in perceptual audio realism, but also in audio-side instruction adherence.
The remaining weakness is concentrated in visible-source grounding. On On-screen, Seedance~2.0 reaches 20.91 in AV Coherence, essentially matching but not exceeding the ground-truth total of 20.93, and remaining well below its other dimension totals. The same internal gap appears for Kling-v3-omni (18.67 in AV Coherence vs.\ 22.29 in Audio Quality) and Veo~3.1 (17.54 vs.\ 22.70). Thus, the updated results refine rather than overturn the original finding: frontier models now produce strong and controllable audio, but synchronizing that audio with specific visible sources remains comparatively difficult.

\noindent\textbf{On-screen Audio Remains the Most Challenging Grounding Regime.}
The On-screen subset simultaneously requires realistic audio, prompt consistency, and precise synchronization with visible sound-producing events. Even Seedance~2.0, which exceeds the On-screen ground-truth aggregate by 5.95 points, remains slightly below the ground-truth AV Coherence total (20.91 vs.\ 20.93). All other evaluated systems fall further behind on this dimension. By comparison, Seedance~2.0 exceeds the ground truth by more than two points in both On-screen Audio Quality and Audio Expressiveness.
This contrast shows that the residual difficulty is not low perceptual realism alone. It lies in binding generated audio to visible speech, instrument playing, and object events. Off-screen evaluation removes explicit visible grounding, but shifts the challenge toward plausible acoustic world-building and faithful control of non-visible sound events. Mixed scenes then test whether these abilities coexist, which Seedance~2.0 handles well in aggregate but not uniformly across all type--dimension cells.

\begin{figure*}[!t]
\centering
\begin{minipage}[t]{0.25\textwidth}
    \centering
    \includegraphics[width=\linewidth,
        trim=8 10 8 7,
        clip]{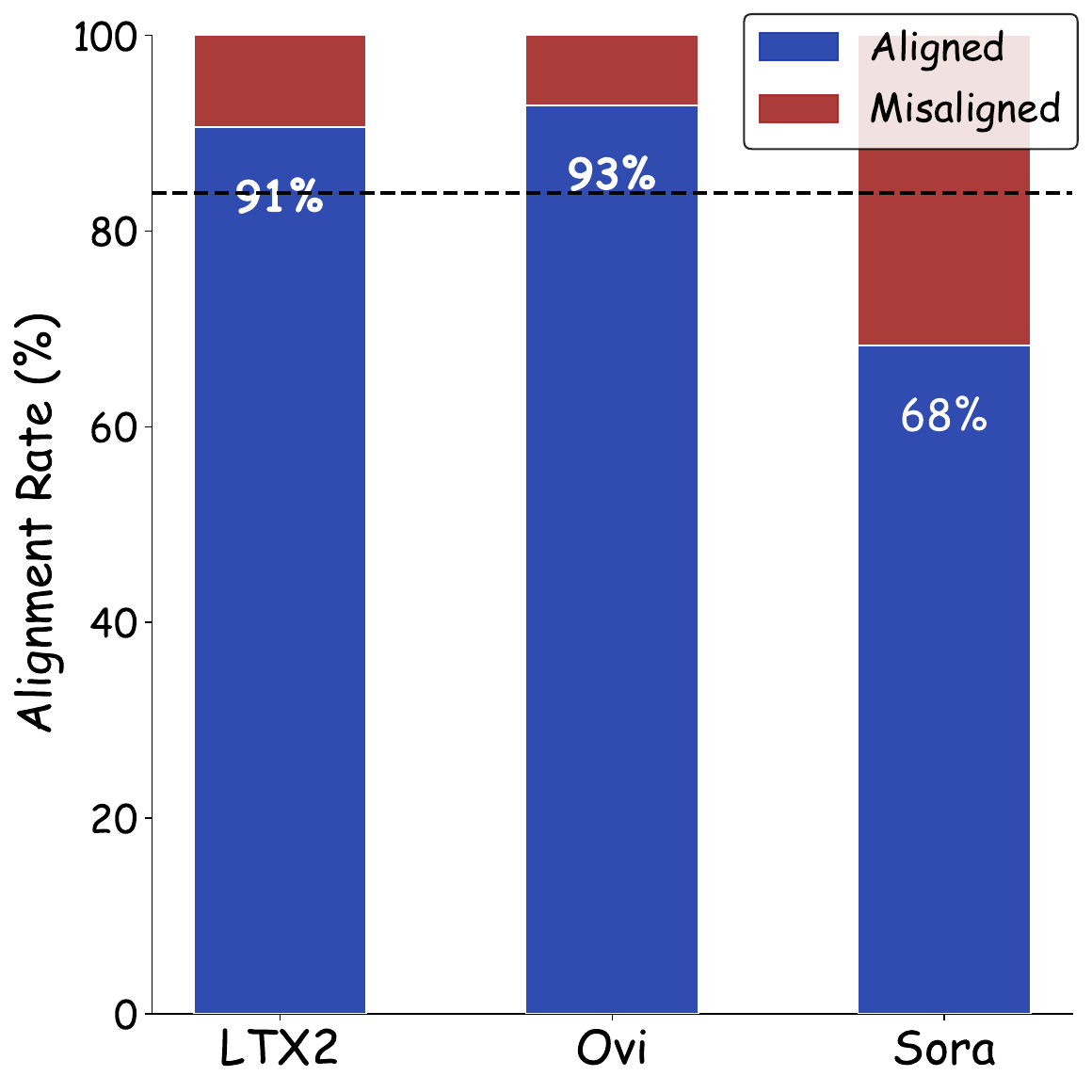}
    {\footnotesize (a) GT vs Model Agreement}
\end{minipage}
\hspace{0.035\textwidth}
\begin{minipage}[t]{0.25\textwidth}
    \centering
    \includegraphics[width=\linewidth,
        trim=8 10 8 7,
        clip]{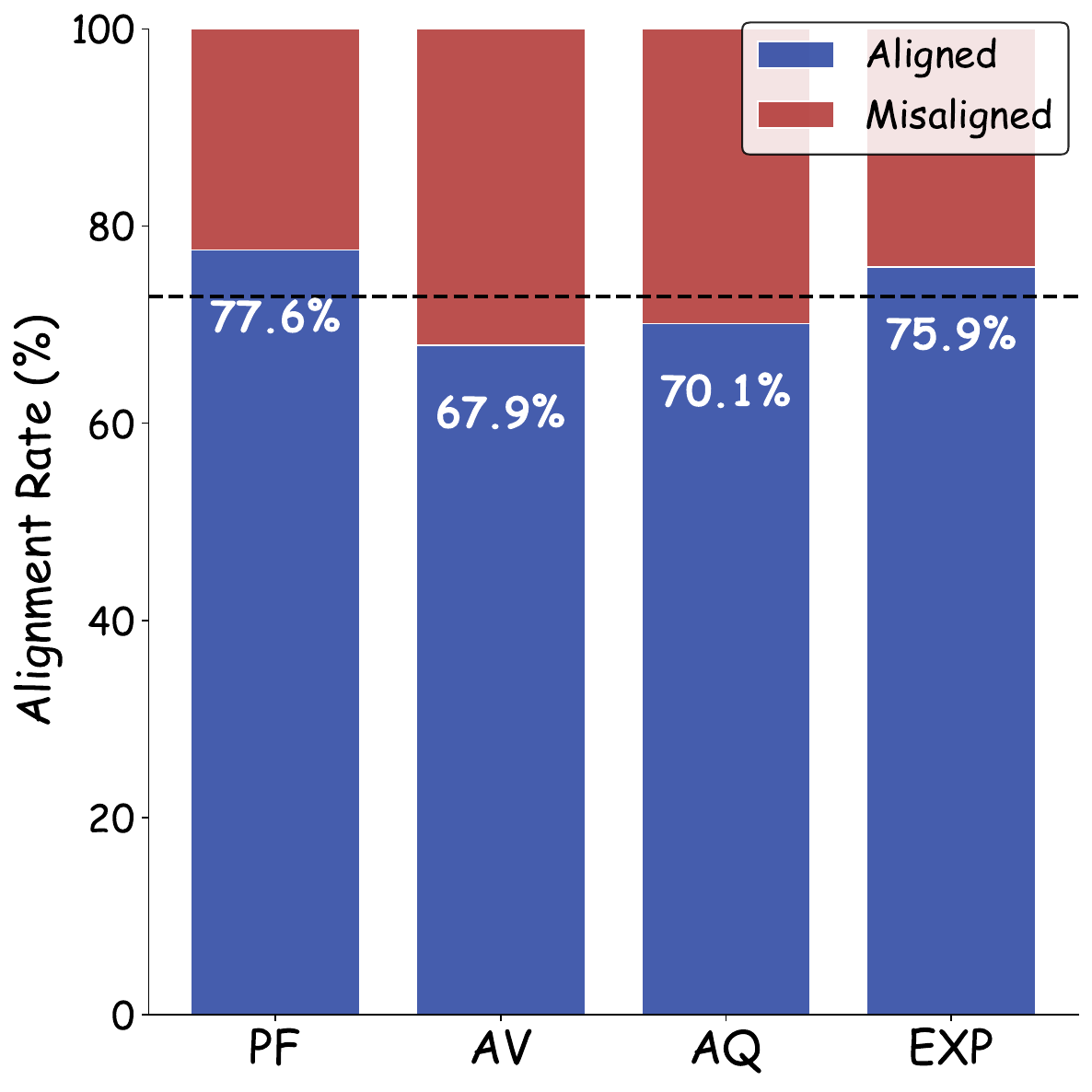}
        
    {\footnotesize (b) Overall Alignment by Dimension}
\end{minipage}
\hspace{0.035\textwidth}
\begin{minipage}[t]{0.25\textwidth}
    \centering
    \includegraphics[width=\linewidth,
        trim=8 10 8 7,
        clip]{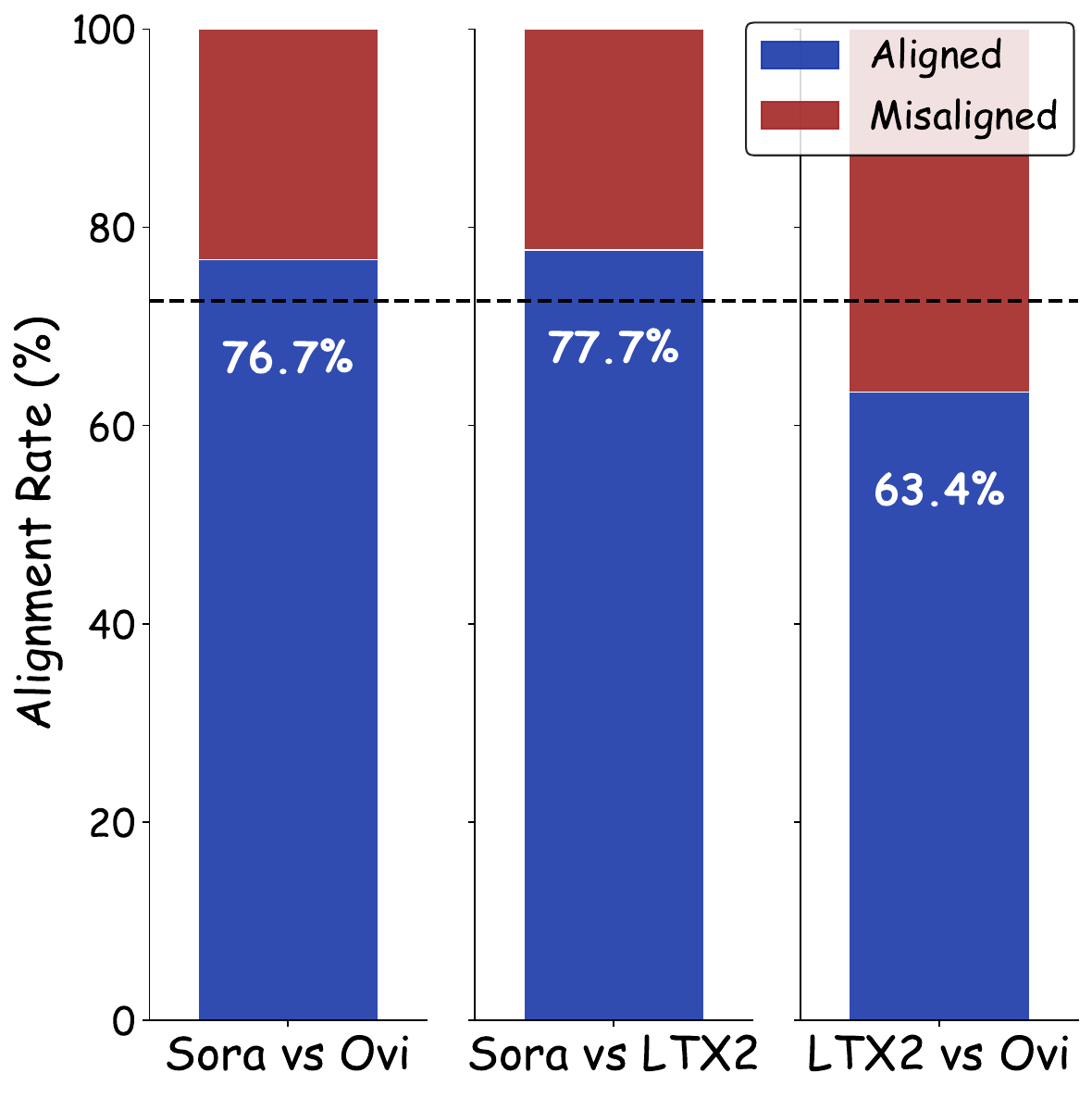}
        
    {\footnotesize (c) Pairwise Alignment Overall}
\end{minipage}
\vspace{ -2 pt}
\caption{Alignment analysis of model-based evaluation against human judgments.}
\label{fig:three_square_top}
\end{figure*}

\noindent\textbf{Music Grounding Improves, but Remains Model-Dependent.}
Seedance~2.0 changes the earlier picture that Music is uniformly the weakest audio type. Its On-screen subtype totals are nearly balanced at 32.62 for Speech, 32.54 for Music, and 32.58 for Sound. Nevertheless, its Music AV Coherence score (6.46) remains lower than its Sound score (7.81), showing that aggregate balance can conceal a grounding-specific weakness.
The weakness is more pronounced for other frontier systems. On Kling-v3-omni, On-screen AV Coherence is 7.75 for Speech, 4.06 for Music, and 6.86 for Sound; on Veo~3.1, the corresponding values are 5.73, 4.77, and 7.03. Their Music subtype totals also remain substantially below Speech. The new model therefore narrows the aggregate Music gap, but visible musical-performance grounding remains a useful stress test that distinguishes capability profiles across models.

\noindent\textbf{PRISM-Bench Reveals Failure Modes Hidden by Aggregate Scores.}
A principal value of \bench{} is that it exposes failure modes that aggregate scores alone would obscure, including for the new leader. Seedance~2.0 achieves an On-screen Final score of 97.73, yet its AV Coherence total is 20.91, compared with 25.24 in Audio Quality, 25.69 in Audio Expressiveness, and 25.89 in Prompt Following. On Off-screen, its Sound total (24.74) trails Speech (26.54) and Music (26.83), with the largest type gap appearing in Prompt Following. Kling-v3-omni exhibits a different profile: its On-screen Speech total is 33.82, but its Music total drops to 23.72.
These examples demonstrate that a strong overall score does not imply uniform behavior across dimensions, audio types, and visibility conditions. By factorizing evaluation along these axes, \bench{} remains informative even when a new model establishes a large aggregate lead: it identifies whether the remaining bottleneck is visible-source binding, musical-performance synchronization, or fine-grained control of Off-screen sound events.

\noindent\textbf{Ground Truth Is Not a Hard Upper Bound.}
The updated results strengthen the observation that the ground-truth reference is an anchor rather than a hard upper bound. Seedance~2.0 exceeds the calibrated ground-truth Final score on On-screen, Off-screen (78.11 vs.\ 71.10), and Mixed (99.88 vs.\ 93.69). The comparison is not uniform across dimensions: on On-screen, Seedance~2.0 exceeds ground truth in Audio Quality, Audio Expressiveness, and Prompt Following, but remains marginally lower in AV Coherence (20.91 vs.\ 20.93); on Mixed, it exceeds the ground-truth total in all four dimensions.
These patterns indicate that the blind, rubric-based protocol does not mechanically favor the reference video. They also clarify how to interpret above-reference performance: generated samples can be judged more perceptually polished or more prompt-adequate than the collected reference without reproducing every reference-specific event. Most importantly, the factorized scores continue to expose residual grounding and control weaknesses after the aggregate score surpasses the reference, so \bench{} does not collapse into a fixed upper-bound test.

\subsection{Human Alignment of Prism-Bench}\label{sec:analysis}

To assess the human alignment of \bench{}, we conduct a study on a representative subset spanning the On-screen, Off-screen, and Mixed settings. We sample outputs from Sora~2, LTX-2, and Ovi. Human raters follow the same factorized protocol as the automated evaluator, assessing Audio-Visual Coherence (AV) and Audio Quality (AQ) from videos alone, and Audio Expressiveness (EX) and Prompt Following (PF) with access to the corresponding captions. Mean agreement exceeds 70\% across dimensions and pairwise comparisons. Detailed annotation procedures and study configurations are provided in the Appendix.

We define \textbf{alignment} as agreement in \textbf{preference direction}: a comparison is aligned when human raters and the MLLM-based evaluator select the same preferred candidate in a Ground Truth--model or model--model comparison; otherwise, it is misaligned. Figure~\ref{fig:three_square_top}(a) reports Ground Truth--model alignment of 91\% for LTX-2, 93\% for Ovi, and 68\% for Sora~2. The lower agreement for Sora~2 is consistent with the greater difficulty of distinguishing stronger generated outputs from the reference when their perceptual gap is smaller.

Figure~\ref{fig:three_square_top}(b) reports alignment across evaluation dimensions. PF achieves the highest agreement at 77.6\%, whereas AV is the most challenging at 67.9\%; the remaining dimensions fall between these values. PF and EX are often supported by explicit semantic or stylistic cues, while AV requires joint assessment of temporal synchronization, visible-source grounding, and scene-dependent plausibility. Mean agreement across dimensions remains above 70\%, supporting the automated protocol as a scalable evaluation proxy.

Figure~\ref{fig:three_square_top}(c) shows agreement of 76.7\%--77.7\% for clearly separated model pairs and 63.4\% for the closer LTX-2 versus Ovi comparison, where preferences are more difficult to distinguish. Taken together, these results support \bench{} as a human-aligned automated proxy for large-scale T2AV evaluation.

\section{Conclusion}\label{sec:conclusion}
We present \bench{}, the first audio-centric diagnostic benchmark for T2AV generation. By organizing evaluation along two orthogonal axes, namely audio type (Speech, Music, and Sound) and sound-source visibility (On-screen and Off-screen), \bench{} provides a more fine-grained and interpretable framework than existing holistic benchmarks. Built on a rigorously curated dataset of around 900 human-verified samples, \bench{} further supports reliable evaluation through a visibility-aware, side-by-side MLLM-as-a-Judge protocol.
Experiments on representative recent T2AV systems show that current progress is uneven across dimensions and audio conditions. While perceptual audio quality has improved substantially, audiovisual grounding remains a major challenge, especially for on-screen music. We hope that \bench{} will facilitate future research on audio-aware multimodal generation and more diagnostic evaluation of T2AV systems.
\begin{acks}
We thank the Meituan LongCat MM team for their essential resource support and the Meituan Visual Intelligence team for their valuable support with data resources. Their contributions greatly facilitated the development and evaluation of PRISM-Bench.

\end{acks}

\balance
\bibliographystyle{ACM-Reference-Format}
\bibliography{references}
\clearpage
\onecolumn
\section*{Appendix}
\appendix
\section{Ground-Truth Reliability Analysis}
\label{app:gt_reliability}

The single GT row reported in the main results table should be interpreted as a context-averaged descriptive reference. Because the same GT clip is repeatedly judged when paired with different candidate models, some reference-side fluctuation across pairing contexts is expected. The key question is therefore not whether such fluctuation exists, but whether it remains small relative to the real performance differences among candidate models. To answer this question, we treat \emph{GT Std}, \(d_{\text{bias}}\), and \(d_{\text{signal}}\) as the primary evidence, and report \(d_{\text{signal}}/d_{\text{bias}}\) only as a compact descriptive summary.

\subsection{Definitions and Subset-Level Summary}
\label{app:gt_stability}

To assess robustness to contextual anchoring, we report two standardized effect sizes. For any pair of candidate models \((m_1,m_2)\), let \(\mathbf{s}_{\mathrm{GT}}^{(m)}\) denote the distribution of GT scores obtained when GT is paired against \(m\) under the same judge and prompt configuration, and let \(\mathbf{s}_{m}\) denote the corresponding distribution of candidate-model totals on the overlapping evaluation items. We define the GT-induced contextual drift as
\begin{equation}
d_{\text{bias}}(m_1,m_2) =
\frac{\left| \mathbb{E}[\mathbf{s}_{\mathrm{GT}}^{(m_1)}] - \mathbb{E}[\mathbf{s}_{\mathrm{GT}}^{(m_2)}] \right|}{s_{\text{pooled}}},
\end{equation}
and the inter-model performance gap as
\begin{equation}
d_{\text{signal}}(m_1,m_2) =
\frac{\left| \mathbb{E}[\mathbf{s}_{m_1}] - \mathbb{E}[\mathbf{s}_{m_2}] \right|}{s_{\text{pooled}}},
\end{equation}
where \(s_{\text{pooled}}\) is the pooled standard deviation of the compared score distributions. For completeness, we also report
\begin{equation}
\mathrm{SNR} = \frac{d_{\text{signal}}}{d_{\text{bias}}},
\end{equation}
as a compact descriptive ratio. All subset-level quantities in Table~\ref{tab:reliability_main} are reported as unweighted averages over all available model pairs. Specifically, \textbf{Std} denotes the average GT standard deviation across comparison contexts, \(d_{\text{bias}}\) and \(d_{\text{signal}}\) denote the average absolute Cohen's \(d\) values computed over all available model pairs, and \textbf{Eval SNR} is the corresponding average ratio \(d_{\text{signal}}/d_{\text{bias}}\). These statistics jointly characterize both the magnitude of context-induced GT drift and its separation from genuine inter-model performance differences at the subset level.

\begin{table}[htbp]
\centering
\caption{\textbf{Subset-level GT stability and bias--signal separation.}}
\label{tab:reliability_main}
\vspace{0.4em}
\small
\begin{tabular}{lcccc}
\toprule
\textbf{Subset} & \textbf{Std} & \textbf{$\mathbf{d_{\text{bias}}}$} & \textbf{$\mathbf{d_{\text{signal}}}$} & \textbf{Eval SNR} \\
\midrule
On-screen  & 1.426 & 0.281 & 0.521 & 1.85 \\
Off-screen & 1.042 & 0.152 & 0.641 & 4.21 \\
Mixed      & 1.237 & 0.233 & 0.814 & 3.49 \\
\bottomrule
\end{tabular}
\end{table}

Table~\ref{tab:reliability_main} shows a clear subset-level pattern. Most importantly, the average model signal remains at least medium in magnitude on all three subsets: \(d_{\text{signal}}=0.521\) for On-screen, \(0.641\) for Off-screen, and \(0.814\) for Mixed. At the same time, the GT-induced contextual drift remains smaller than the model-level signal throughout. The cleanest regime is Off-screen, where \(d_{\text{bias}}=0.152\) remains small while \(d_{\text{signal}}=0.641\) is substantially larger. Mixed is similarly well separated, with \(d_{\text{bias}}=0.233\) versus \(d_{\text{signal}}=0.814\). On-screen is the hardest regime: its \(d_{\text{bias}}\) rises to \(0.281\), reflecting greater sensitivity to visible grounding and synchronization, but its average signal still reaches \(0.521\), so the subset remains informative for comparative diagnosis.

The raw GT fluctuations in Table~\ref{tab:reliability_main} should therefore be read as dispersion indicators rather than as the benchmark decision criterion itself. Their ordering is nevertheless meaningful and consistent with the subset design: On-screen is the most variable, Off-screen the most stable, and Mixed lies between the two.

\subsection{Dimension-Wise Stability Breakdown}
\label{app:dim_stability}

Table~\ref{tab:appendix_dim_stability} localizes the main source of instability. The dominant contribution comes from the synchronization-sensitive branch of \emph{Audio-Visual Coherence}, especially for On-screen speech and music, which reach \(2.145\) and \(2.314\), respectively. This indicates that visible grounding is the most context-sensitive component of the evaluation: once the judge must resolve lip-sync, performer--sound consistency, or event-level onset alignment, small contextual shifts propagate more strongly into the final score.

\begin{table}[htbp]
\centering
\caption{\textbf{Dimension-wise GT stability (average standard deviation) across subsets. ``/'' indicates that the dimension is inactive for that subset.}}
\label{tab:appendix_dim_stability}
\vspace{0.4em}
\small
\setlength{\tabcolsep}{4pt}
\renewcommand{\arraystretch}{0.96}
\begin{tabular}{llccc}
\toprule
\textbf{Dimension} & \textbf{Subtype} & \textbf{On-screen} & \textbf{Off-screen} & \textbf{Mixed} \\
\midrule
\multirow{3}{*}{Audio-Visual Coherence}
& Speech & 2.145 & /     & 1.929 \\
& Music  & 2.314 & /     & 0.944 \\
& Sound  & 1.718 & /     & 1.308 \\
\cmidrule(lr){1-5}
\multirow{3}{*}{Audio Quality}
& Speech & 1.283 & 0.801 & 1.163 \\
& Music  & 1.465 & 0.889 & 0.925 \\
& Sound  & 1.277 & 1.054 & 1.057 \\
\cmidrule(lr){1-5}
\multirow{3}{*}{Audio Expressiveness}
& Speech & 0.964 & 0.881 & 1.021 \\
& Music  & 1.154 & 0.990 & 1.090 \\
& Sound  & 1.177 & 1.155 & 1.241 \\
\cmidrule(lr){1-5}
\multirow{3}{*}{Prompt Following}
& Speech & 0.966 & 0.845 & 1.041 \\
& Music  & 1.287 & 1.211 & 1.405 \\
& Sound  & 1.395 & 1.571 & 1.497 \\
\bottomrule
\end{tabular}
\end{table}

By contrast, the Off-screen subset removes \emph{Audio-Visual Coherence} entirely, and the remaining branches are generally more stable: several entries fall below \(1.0\), including Off-screen \emph{Audio Quality} for speech (\(0.801\)) and music (\(0.889\)), as well as Off-screen \emph{Audio Expressiveness} for speech (\(0.881\)). The Mixed subset behaves as an intermediate regime. It retains an \emph{Audio-Visual Coherence} branch, but only for the on-screen portion of the active tags, which reduces music-side AV variance relative to pure On-screen while preserving a more difficult speech-side grounding problem. In short, the higher subset-level fluctuation of On-screen is driven primarily by synchronization-sensitive dimensions rather than by a uniform instability across the entire rubric.

\subsection{Interpretation of Bias, Signal, and SNR}
\label{app:bias_signal}

Our main validity criterion is the joint behavior of \(d_{\text{bias}}\) and \(d_{\text{signal}}\): contextual drift should remain limited, while model-level separation should remain at least moderate. Under this criterion, the benchmark remains reliable across all three subsets. Off-screen and Mixed provide the clearest margins, as they combine relatively small \(d_{\text{bias}}\) with strong \(d_{\text{signal}}\). On-screen is more conservative, but its elevated bias is concentrated in the visible grounding branch identified in Table~\ref{tab:appendix_dim_stability}, rather than indicating a general collapse of evaluator consistency.

SNR is therefore reported as a compact descriptive summary rather than as the sole decision criterion. This is particularly appropriate for synchronization-heavy settings, where the ratio can be depressed by harder contextual conditions or near-tie comparisons even when the evaluator still preserves meaningful model separation. This is exactly the case for the On-screen subset: although its average \(\mathrm{SNR}=1.85\), its average \(d_{\text{signal}}=0.521\) remains in the medium-effect range, so the evaluator continues to distinguish models by a non-trivial margin. The practical implication is visibility-aware: Off-screen and Mixed provide the cleanest reliability margins, while On-screen remains informative but should be interpreted more cautiously for closely matched systems.

\section{Judge Model Ablation}
\label{app:judge_ablation}

We further conduct a judge-model ablation under a controlled evaluation setup. All judge variants are evaluated on the same Mixed-set protocol, with the same two-stage prompt structure, the same blind two-candidate input format, and the same JSON output schema. We use the Mixed subset because it jointly includes on-screen and off-screen conditions, making it the most comprehensive setting for judge selection. The comparison thus isolates judge behavior rather than differences in prompt engineering or interface design. The purpose is not merely to identify the judge with the lowest variance, but to examine whether a candidate judge can maintain limited contextual drift while still preserving meaningful rubric-aligned discrimination. Table~\ref{tab:appendix_judge_ablation} summarizes the resulting reliability statistics.

\begin{table}[htbp]
\centering
\caption{\textbf{Judge-model ablation under a controlled evaluation setup. \textbf{Std} is the average GT standard deviation, and \(d_{\text{bias}}\), \(d_{\text{signal}}\), and \textbf{Eval SNR} are computed in the same way as in Table~\ref{tab:reliability_main}.}}
\label{tab:appendix_judge_ablation}
\vspace{0.4em}
\small
\begin{tabular}{lcccc}
\toprule
\textbf{Judge} & \textbf{Std} & \textbf{$\mathbf{d_{\text{bias}}}$} & \textbf{$\mathbf{d_{\text{signal}}}$} & \textbf{Eval SNR} \\
\midrule
\textbf{Gemini-3.1 Pro}  (Final) & 0.915 & 0.141 & 0.429 & 3.05 \\
Gemini-3.1 Flash              & 0.524 & 0.075 & 0.273 & 3.64 \\
Gemini-3.0 Flash              & 0.591 & 0.073 & 0.367 & 5.00 \\
Qwen3-Omni-30B-Instruct                          & 0.167 & 0.011 & 0.012 & 1.11 \\
\bottomrule
\end{tabular}
\end{table}

Quantitative Comparison and Final Selection. The ablation suggests an important selection principle: an extremely small \(d_{\text{bias}}\) is not necessarily optimal by itself. A desirable judge should keep bias low while still preserving sufficient score variation to separate models with genuinely different performance. This trade-off is illustrated by \textbf{Qwen3-Omni}. It achieves the smallest \(d_{\text{bias}}\), indicating strong scoring consistency. However, its \(d_{\text{signal}}\) is similarly small, resulting in an \textbf{Eval SNR} of only 1.11. This pattern suggests that, under our evaluation setup, \textbf{Qwen3-Omni} uses the score scale rather conservatively, so its stability advantage does not translate into equally strong discriminative power. One possible explanation is that our blind pairwise video-evaluation protocol may not be well aligned with the input format or supervision signals emphasized during its training. We therefore interpret this result not as a general weakness of Qwen3-Omni, but as evidence that judge selection should balance low bias with sufficient sensitivity to meaningful quality differences.

The two flash variants show a milder version of the same trade-off. Both \textbf{Gemini-3.1 Flash} and \textbf{Gemini-3.0 Flash} reduce contextual drift to around \(0.07\)\,--\,\(0.08\), but they also reduce the average model-level signal to \(0.273\) and \(0.367\), respectively. These values indicate that the judges are comparatively conservative and may compress meaningful score differences, especially when the compared models are not extremely far apart.

By contrast, \textbf{Gemini-3.1 Pro} maintains a small but non-trivial contextual bias (\(d_{\text{bias}}=0.141\)), while preserving the largest average inter-model separation among the tested judges (\(d_{\text{signal}}=0.429\)). This balance is preferable for \bench. The selected judge should not drive \(d_{\text{bias}}\) toward zero by becoming insensitive; rather, it should retain enough rubric responsiveness to separate genuinely different candidates while keeping bias materially below signal. Under this criterion, \textbf{Gemini-3.1 Pro} provides the most reasonable operating point.

A more concrete indication appears in the pairwise ablation results. Under \textbf{Gemini-3.1 Pro}, the strongest model contrast reaches \(d_{\text{signal}}=0.639\), which is already in the medium-to-large range, while the corresponding average \(d_{\text{bias}}\) remains well below that level. This pattern is exactly what the benchmark needs: small contextual drift together with non-compressed model separation. The final judge is therefore chosen not because it minimizes every reliability statistic in isolation, but because it best preserves the intended trade-off between stability and discriminability.

\section{Evaluation Protocol Details}
\label{app:protocol}
The implementation of the visibility-aware judging protocol in \bench is described below. All three subsets are evaluated under a common blind side-by-side scaffold, while differing only in how visibility assumptions determine the dimensions that are meaningfully instantiated. Figures~\ref{fig:appendix_prompt_shared} and~\ref{fig:appendix_prompt_stagewise} illustrate the shared scaffold and its stage-wise slots, respectively. In particular, the system-level instruction, blind candidate setup, preprocessing rules, score scale, and response schema are held fixed across subsets. The only variation lies in how subset-specific visibility conditions govern dimension activation and tag routing. Moreover, the reference prompt is not a separate prompt family component, but a shared field that is populated only in the caption-conditioned stage.

\subsection{Subset-Specific Judge Configuration}
\label{app:subset_prompt}

\begin{figure}[htbp] 
\centering \vspace*{-2.5em} \makebox[\textwidth][c]{ \includegraphics[width=1.2\textwidth,trim=1 140 1 10,clip]{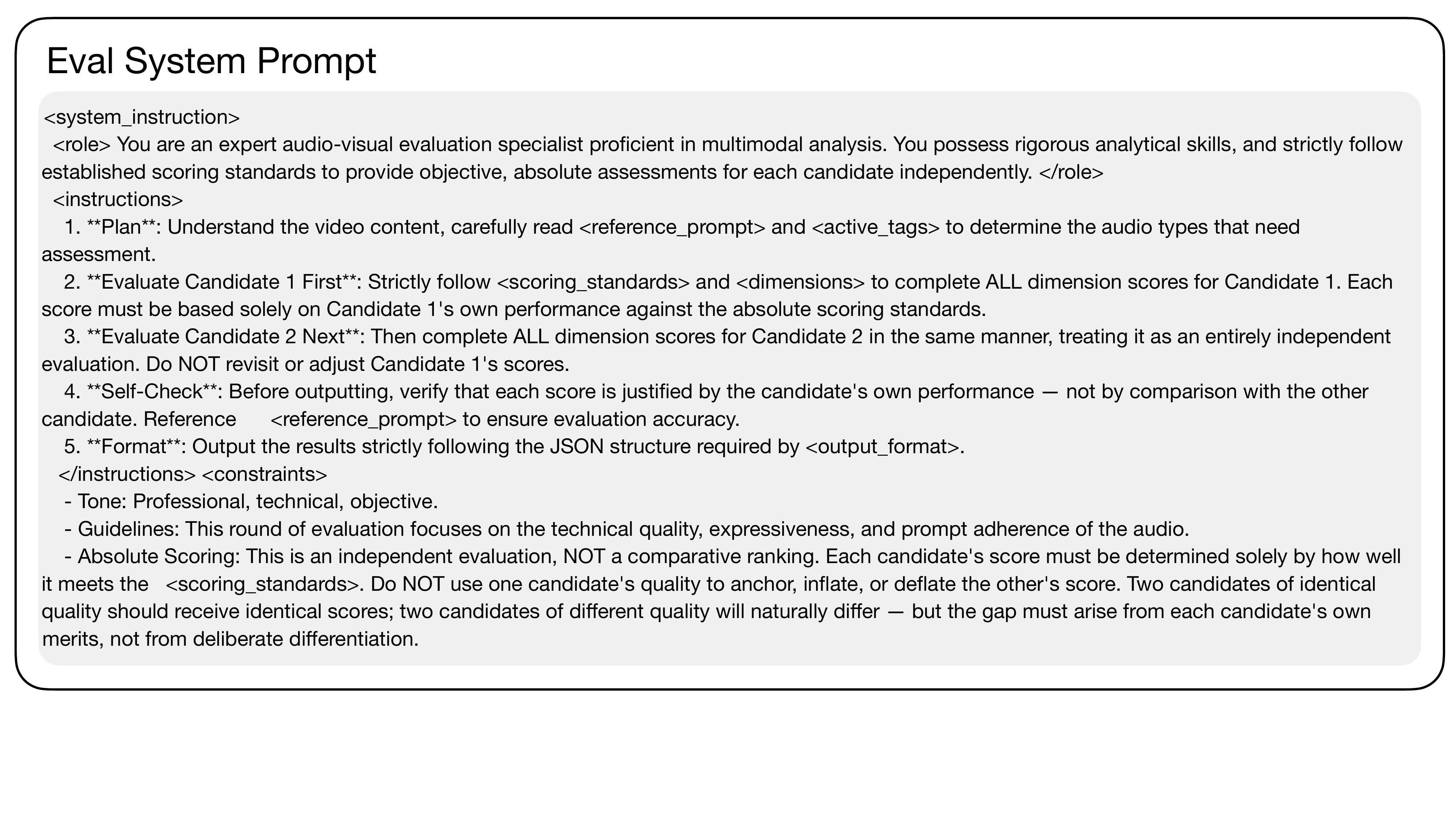} } \vspace{-0.3em} \makebox[\textwidth][c]{ \includegraphics[width=1.2\textwidth,trim=1 90 1 20,clip]{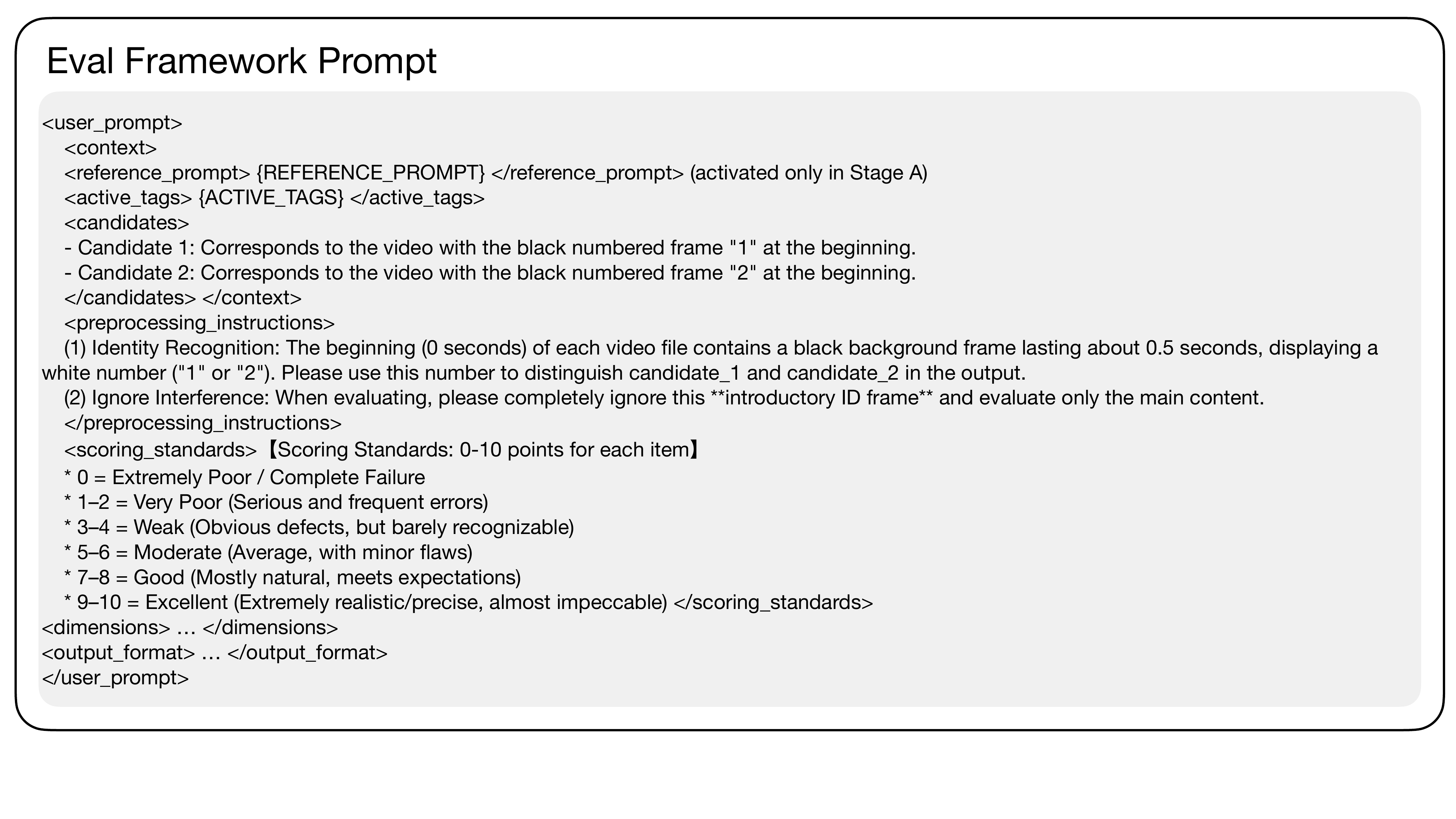} } \caption{\textbf{Shared prompting scaffold of the judge model.}} \label{fig:appendix_prompt_shared} 
\end{figure}

\begin{figure}[htbp] \centering \vspace*{-2.8em} \makebox[\textwidth][c]{ \includegraphics[width=1.2\textwidth,trim=1 20 1 1,clip]{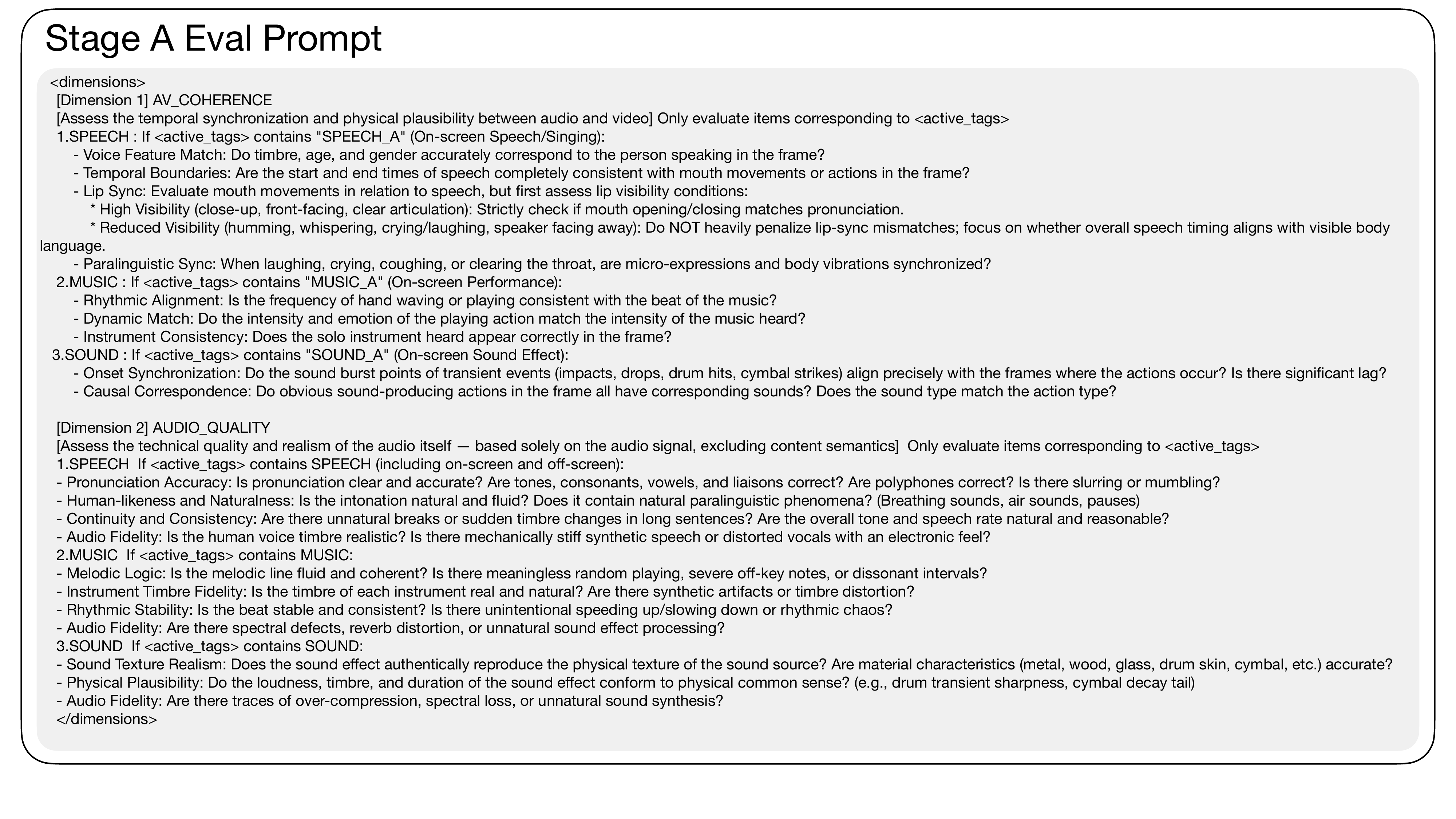} } \vspace{-0.05em} \makebox[\textwidth][c]{ \includegraphics[width=1.2\textwidth,trim=1 2 1 1,clip]{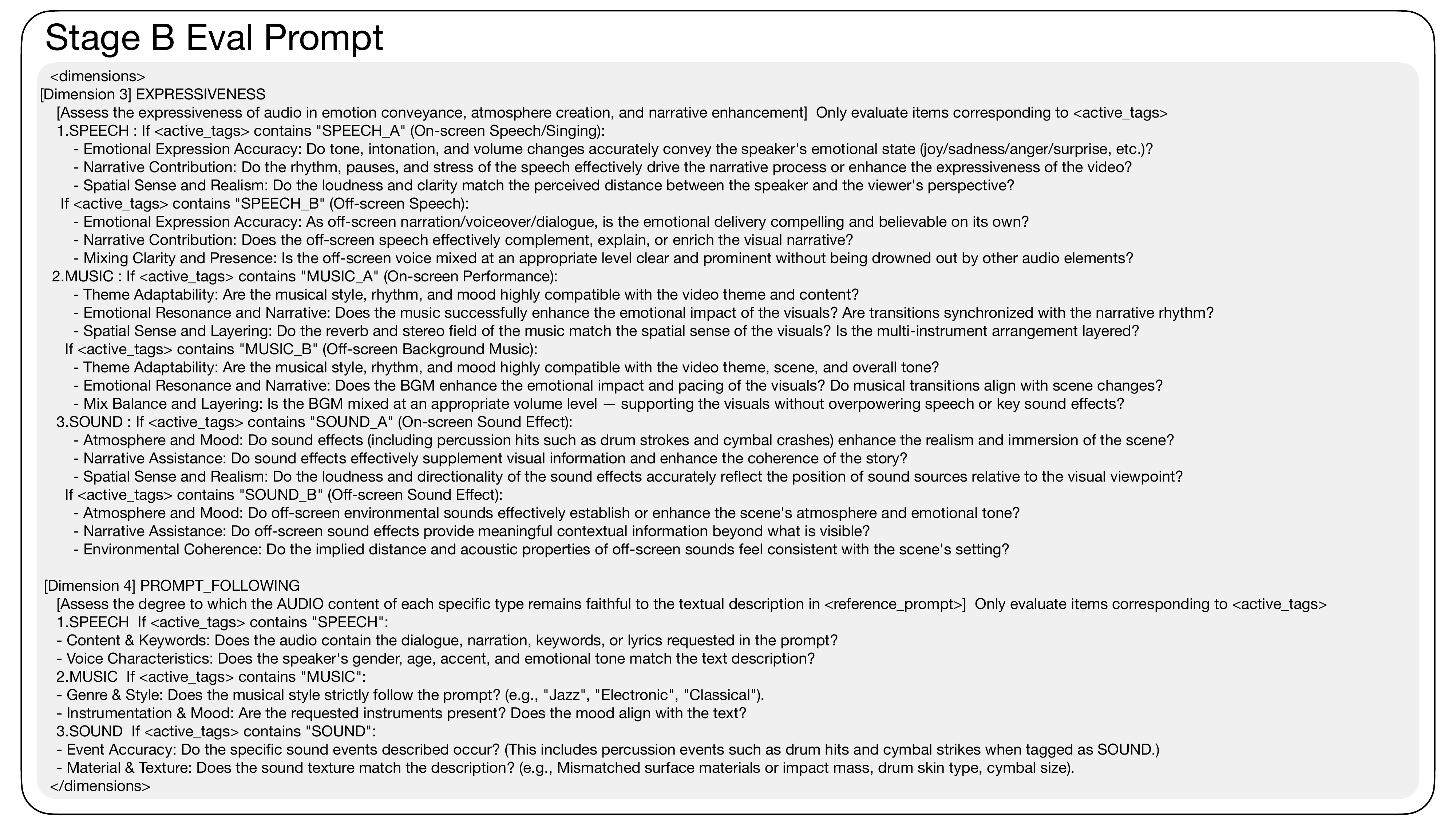} } \caption{\textbf{Stage-specific prompt slots in the shared judging scaffold.}} \label{fig:appendix_prompt_stagewise} \end{figure}

The protocol follows a shared two-stage formulation, illustrated in Figures~\ref{fig:appendix_prompt_shared} and~\ref{fig:appendix_prompt_stagewise}, so that differences across subsets reflect diagnostic intent rather than prompt-format variation. As shown in Figure~\ref{fig:appendix_prompt_shared}, all subsets inherit the same global scaffold, including the system-level instruction, blind candidate setup, preprocessing rules, scoring scale, and response schema. Figure~\ref{fig:appendix_prompt_stagewise} further shows how this scaffold is instantiated through two stage-specific prompt slots. Stage~A focuses on perceptual grounding without textual conditioning and is used to assess dimensions whose validity depends primarily on direct audio--visual evidence. Stage~B introduces the human-verified caption and evaluates dimensions that require semantic interpretation of the generated audio relative to the intended scene description. This separation keeps the judging process structurally consistent while preventing prompt information from leaking into dimensions that should be determined from perceptual evidence alone.

Within this common scaffold, subset-specific behavior is determined by the visibility assumption encoded in the active tags. For the \emph{On-screen} subset, the protocol assumes that the relevant sound-producing source is visually observable, so synchronization and source-grounded correspondence remain valid evaluation targets. Under this setting, Stage~A assesses \emph{AV Coherence} together with \emph{Audio Quality}, while Stage~B evaluates \emph{Expressiveness} and \emph{Prompt Following} with caption conditioning. For the \emph{Off-screen} subset, by contrast, the protocol explicitly removes visible-source correspondence from the decision space: since the causally relevant source is not expected to appear in the frame, the evaluation should not penalize the absence of on-screen alignment. As a result, \emph{AV Coherence} is deactivated, whereas the remaining dimensions continue to be assessed under an off-screen interpretation.

The \emph{Mixed} subset inherits the same global structure but requires visibility-aware routing within a single request, because on-screen and off-screen audio conditions may coexist in the same sample. In this case, only the on-screen portion of the active tags contributes to \emph{AV Coherence}, since strict source-grounded alignment is meaningful only when the sound source is visible. By contrast, \emph{Audio Quality} remains applicable to all active audio types regardless of visibility, and Stage~B evaluates \emph{Expressiveness} and \emph{Prompt Following} over the full tag set with caption conditioning. The mixed setting therefore does not introduce a new prompt template; instead, it uses the shared scaffold in Figure~\ref{fig:appendix_prompt_stagewise} to selectively activate visibility-sensitive criteria while preserving a unified response format across heterogeneous audio conditions.

At the request level, each judge call is organized as a single multimodal message consisting of a text instruction followed by two blinded candidate videos. In the evaluation implementation, the two videos are uploaded separately as Base64-encoded video parts rather than pre-concatenated into one merged clip. Candidate identity is recovered from the 0.5-second black identifier frame at the beginning of each clip, labeled ``1'' or ``2'', and that introductory frame is explicitly ignored during scoring.

\begin{figure}[htbp]
    \centering
    \vspace{-0.10em}
    \makebox[\textwidth][c]{
    \includegraphics[width=1.2\textwidth,trim=1 20 1 1,clip]{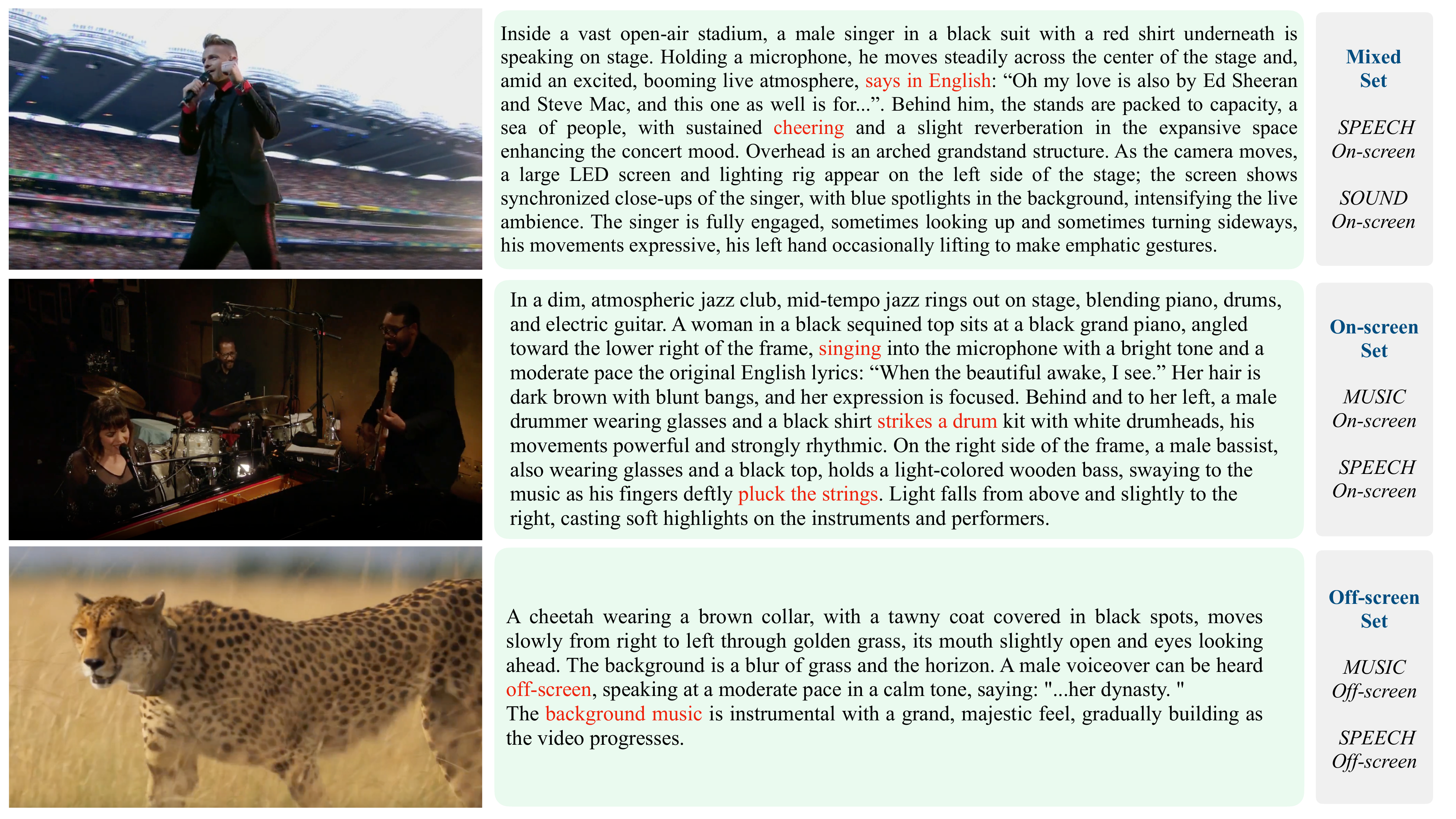}
    }
    \caption{\textbf{Caption examples and subset-aware tag routing. }}
    \label{fig:appendix_caption_examples}
    \vspace{-0.10em}
\end{figure}

For the On-screen and Mixed subsets, the implementation issues two judge calls per sample. Stage~A receives the active tags and the two candidate videos, and returns scores and evidence for \emph{AV Coherence} and \emph{Audio Quality}. Stage~B additionally receives the human-verified caption as the reference prompt and returns scores and evidence for \emph{Expressiveness} and \emph{Prompt Following}. For the Off-screen subset, the same blind input structure is preserved, but the active dimensions are reduced because \emph{AV Coherence} is inactive.

The role of the caption is therefore not merely descriptive. It determines which semantic branch is activated in the caption-conditioned stage and clarifies how the same audio-type tags are interpreted under the On-screen, Off-screen, and Mixed protocols. Representative examples of subset-specific captions and their activated tags are illustrated in Figure~\ref{fig:appendix_caption_examples}.

\subsection{\textbf{Case Study: Stage-Wise Score Aggregation}}
\label{app:case_study}

\begin{figure}[htbp]
    \centering
    \vspace*{-5.0em}
    \makebox[\textwidth][c]{
      \includegraphics[width=1.06\textwidth,trim=1 1 1 5,clip]{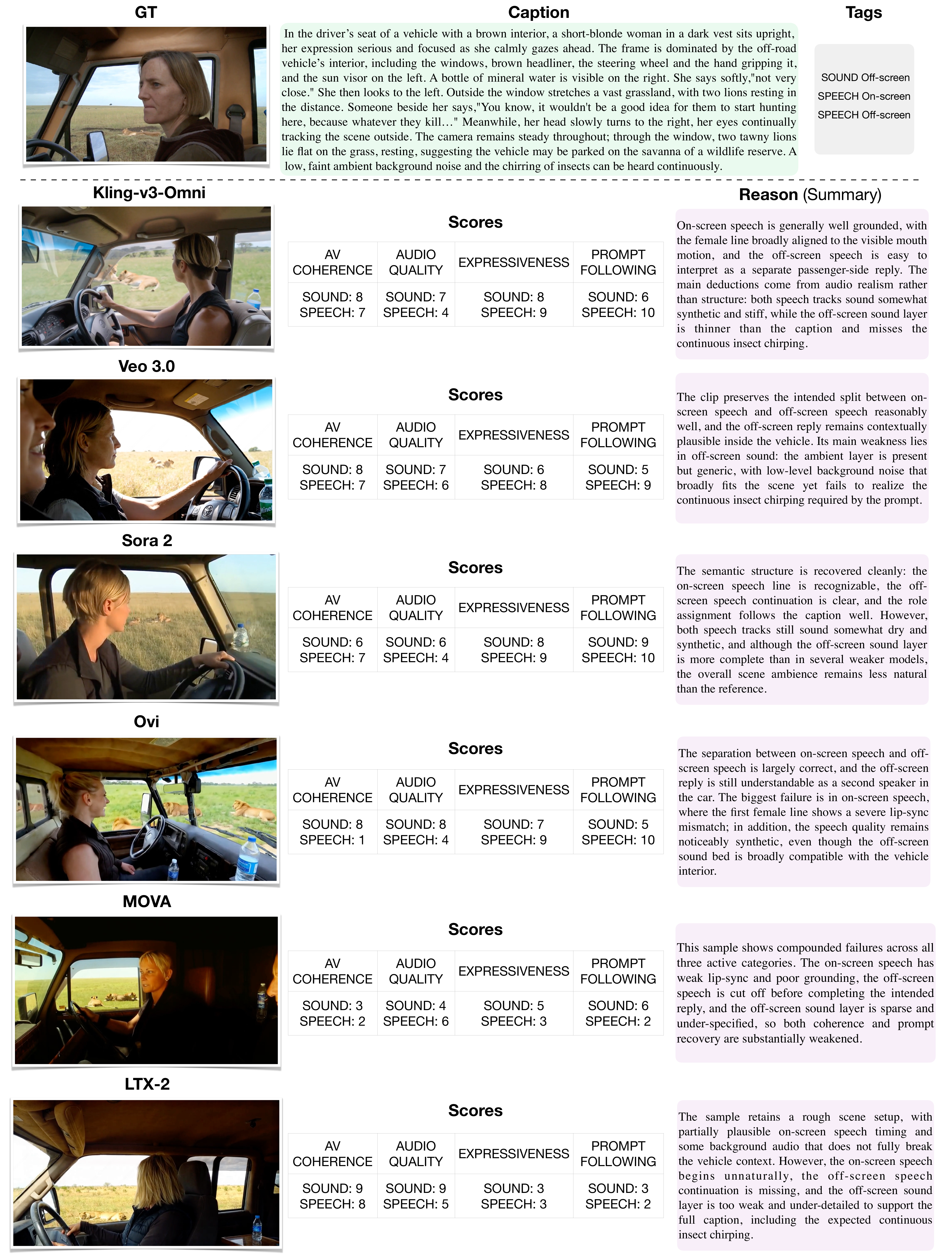}
    }
    \vspace*{-2.0em}
    \caption{\textbf{Suggested case study for stage-wise score aggregation.}}
    \label{fig:appendix_case_study}
\end{figure}

After both stages succeed, the implementation merges the two JSON responses candidate by candidate. The final record copies \emph{AV Coherence} and \emph{Audio Quality} from Stage~A, copies \emph{Expressiveness} and \emph{Prompt Following} from Stage~B, and then verifies that all required dimensions are present before saving the unified evaluation record. In this sense, the final output is not produced by a third scoring pass; it is a deterministic merge of the two stage-wise results.

Figure~\ref{fig:appendix_case_study} provides a concrete case study of this process. Rather than emphasizing the activation procedure itself, the figure is intended to show how the protocol operates on an individual sample: one caption, its activated tags, the stage-wise score objects returned for the two candidates, and the final merged record. This sample-level view makes the evaluation outcome directly auditable and offers a more informative illustration of judging validity than a purely schematic request-format diagram.

\section{Human Alignment}
\label{app:human_alignment}
Following common practice in human evaluation, each comparison pair is independently rated by \textbf{three human annotators}. All annotators assign absolute integer scores on a \textbf{0--10 scale} using the same rubric as the automated judge. For each scoring item, the final human score is reported as a single integer label obtained by majority vote. In the rare case where all three ratings differ, the item is adjudicated through a brief reconciliation step to determine the final integer score. This protocol is intended to reduce individual rater noise while preserving a discrete and interpretable human reference.

The human study covers outputs from three representative generation models, \textbf{Sora~2}, \textbf{LTX-2}, and \textbf{Ovi}. We select these models because they span different model families and performance levels in the main benchmark, allowing human alignment to be examined across a broader range of generation quality rather than within a single performance tier.

Pair allocation is stratified by valid tag combinations rather than by random subset-level sampling. For each subset, we first enumerate the valid combinations of active tags and then sample four examples for each combination. The On-screen subset contains seven valid tag combinations, resulting in 28 pairs in total. The Off-screen subset follows the same design and also contains seven valid tag combinations, resulting in 28 pairs. The Mixed subset allows on-screen and off-screen tags to co-occur within the same sample and therefore contains 15 valid tag combinations, resulting in 60 pairs. This design ensures coverage of the subset-specific tag space without overrepresenting frequent patterns.

Each task is presented as a blind GT-versus-model comparison. The two clips are assigned randomized identities, and each begins with a 0.5-second black frame containing a white numeric identifier. Annotators use that identifier only to map scores to Candidate~1 or Candidate~2; the identifier frame itself is ignored during quality judgment. As in the automatic pipeline, inactive audio types are not scored.

The manual protocol mirrors the automated two-stage design. In Stage~1, raters watch the paired clips without reading the prompt and score only \emph{AV Coherence} and \emph{Audio Quality}. \emph{AV Coherence} is restricted to on-screen audio and focuses on timing, causal matching, and visible-source consistency. \emph{Audio Quality} assesses the realism and technical fidelity of speech, music, and sound regardless of meaning. In Stage~2, raters read the prompt and then score \emph{Expressiveness} and \emph{Prompt Following}, again only for the audio types that are present in the current sample. The score interpretation is identical to that used by the judge model: 9--10 for excellent, 7--8 for good, 5--6 for moderate, 3--4 for weak, 1--2 for very poor, and 0 for complete failure.

\begin{figure}[htbp]
    \centering
    \makebox[\textwidth][c]{
      \includegraphics[width=1.1\textwidth,trim=10 380 10 5,clip]{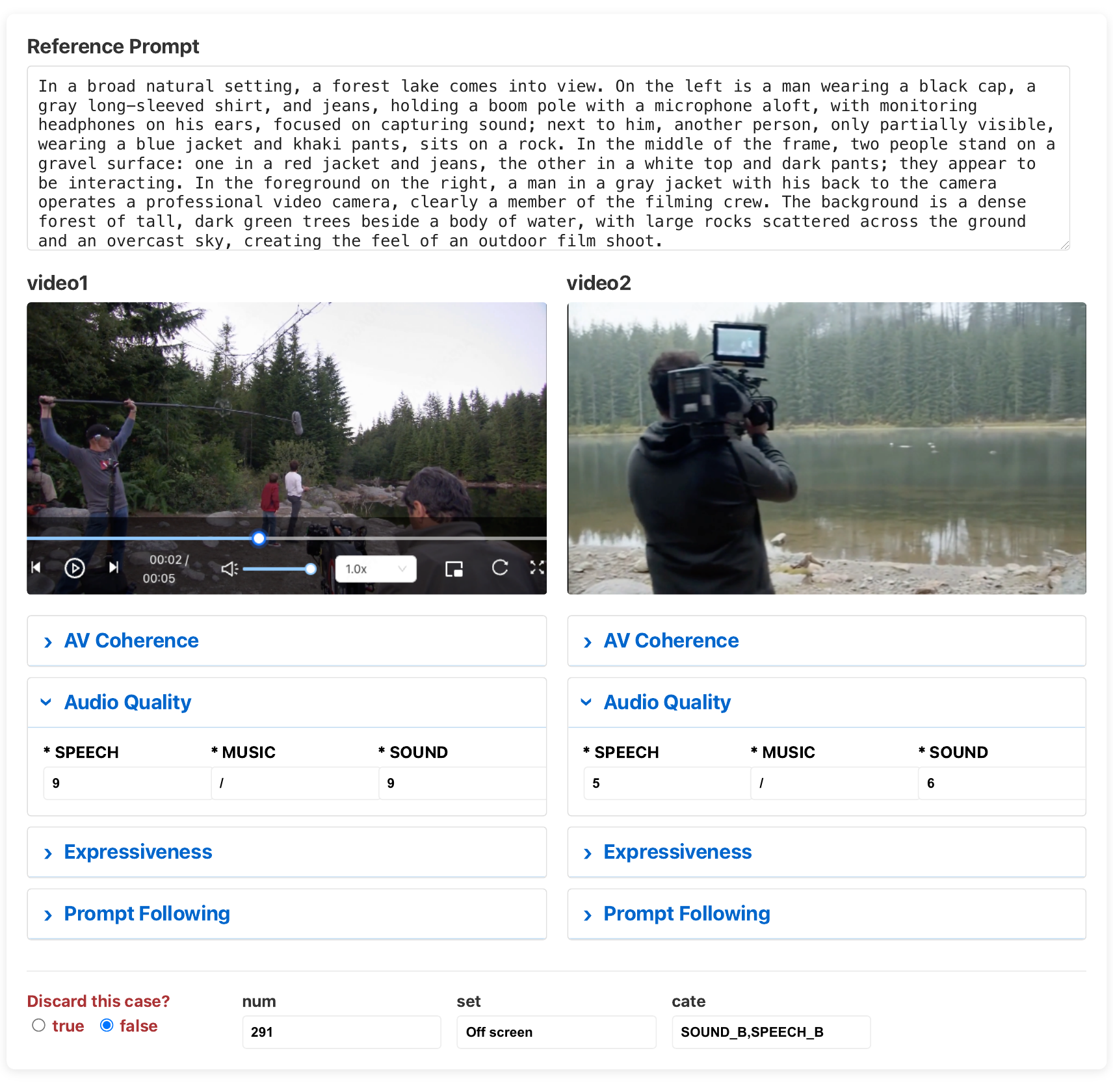}
    }
    \caption{\textbf{Human evaluation interface used in the alignment study.}}
    \label{fig:appendix_human_ui}
\end{figure}

Figure~\ref{fig:appendix_human_ui} shows the human annotation interface used in the alignment study. The interface presents the English prompt together with two blinded candidate videos displayed side by side, and organizes the four evaluation dimensions into stage-specific scoring panels. It also records sample metadata such as \textit{num}, subset, and active tag category, while providing a discard option for invalid cases.

\end{document}